# Brain-wide 3D imaging of neuronal activity in *Caenorhabditis elegans* with sculpted light

Tina Schrödel[1,5], Robert Prevedel[1,2,3,5], Karin Aumayr[1,4], Manuel Zimmer[1] & Alipasha Vaziri[1,2,3]

[1] Research Institute of Molecular Pathology, Vienna, Austria
[2] Max F. Perutz Laboratories, University of Vienna, Vienna, Austria
[3] Research Platform Quantum Phenomena & Nanoscale Biological Systems (QuNaBioS), University of Vienna, Vienna, Austria
[4] Institute of Molecular Biotechnology, Austrian Academy of Sciences, Vienna, Austria

[5] These authors contributed equally to this work.

Correspondence should be addressed to M.Z. (zimmer@imp.ac.at) or A.V. (vaziri@imp.ac.at).

**Recent efforts in neuroscience research seek to obtain detailed anatomical neuronal wiring maps as well as information on how neurons in these networks engage in dynamic activities. Although the entire connectivity map of the nervous system of *C. elegans* has been known for more than 25 years, this knowledge has not been sufficient to predict all functional connections underlying behavior. To approach this goal, we developed a two-photon technique for brain-wide calcium imaging in *C. elegans* using wide-field temporal focusing (WF-TEFO). Pivotal to our results was the use of a nuclear-localized, genetically encoded calcium indicator (NLS-GCaMP5K) that permits unambiguous discrimination of individual neurons within the densely-packed head ganglia of *C. elegans*. We demonstrate near-simultaneous recording of activity of up to 70% of all head neurons. In combination with a lab-on-a-chip device for stimulus delivery, this method provides an enabling platform for establishing functional maps of neuronal networks**

One of the fundamental goals of neuroscience is to understand how external sensory inputs and internal states are represented and processed by neuronal circuits in the nervous system to generate behavior. While in both simple and more complex organisms there is evidence of the existence of discrete anatomically pathways that compute sensory input and generate motor output, there is an increasing understanding that, most sensory functions and behavioral states are represented in a flexible and distributed fashion across large neuronal networks [1, 2]. An understanding of the underlying mechanisms of how brains compute information and generate behaviors requires both detailed anatomical information on connectivity [3, 4] and information on how the involved neuronal populations engage in functionally relevant dynamic activity patterns [5, 6]. Neuroscience research is making progress along both directions, facilitated by technological development. Ideally, both the anatomical and the functional mapping should be done in a brain-wide fashion, but the complexity of nervous systems and the vast number of neurons is still an impediment towards this goal. In this respect, impressive results using $Ca^{2+}$-imaging have recently been obtained in larval zebrafish brains [7, 8].

The simple nematode model organism *C. elegans,* with its 302 neurons and 8000 synaptic connections, is the only animal for which a complete nervous system has been anatomically mapped [9]. Therefore, it provides a unique opportunity to establish a complete map of structure-function relationships in an entire nervous system for the first time. However, current research in *C. elegans* focuses mainly on isolated neurons and small circuits. This is mainly due to technical limitations in electrophysiology and functional imaging: *C. elegans* neurons are small and densely packed into head and tail ganglia and closely surrounded by neuropil, which makes them inaccessible to multi-electrode arrays and hard to tract using image segmentation methods when using pan-neuronally expressed genetically encoded $Ca^{2+}$-indicators. To overcome these limitations, a high resolution imaging technique capable of unbiased, fast volumetric imaging of neurons as well as $Ca^{2+}$-reporters that allow unambiguous anatomical identification of neurons and their $Ca^{2+}$-signals would be required.

While various approaches to speed up image acquisition rate in functional imaging experiments have been reported, including capturing the activity of selected neurons [10-12], until very recently [7, 8, 13] the available imaging techniques were not capable of unbiased fast volumetric readout of activity in large neuronal populations at single cell resolution in any model organism. One of the reasons for this has been that conventional optical excitation setups exhibit an inherent trade-off between axial resolution and lateral excitation area. But several new excitation schemes that retain high axial resolution while providing wide-field excitation are now available [13-20].

Here, we introduce wide-field temporal focusing (WF-TEFO), a two-photon technique, which is based on light-sculpting and enables recording, with high temporal and spatial resolution, the activity of the majority of neurons in the head ganglia of *C. elegans*. To aid in cell segmentation, we use transgenic worms expressing the $Ca^{2+}$-sensor GCaMP5K in a pan-neuronal and nucleus-bound fashion. We also used custom-designed microfluidic devices to restrain the worms while applying chemosensory stimuli. Using this approach, we demonstrate brain-wide and near-simultaneous $Ca^{2+}$- imaging of 70% of the neurons contained in the head ganglia. The spatial and temporal resolution of our imaging technique is adequate for investigating global properties of the *C. elegans* nervous system, such as correlated activity among groups of neurons and their responses to sensory stimuli. This paves the way for future investigations on how sensory information is processed at the level of the whole brain, and for establishing a functional map of *C. elegans'* nervous system and potentially also in other model organisms.

## RESULTS

### Volumetric imaging using WF-TEFO

Our approach for volumetric imaging is based on temporal focusing [15-18, 20, 21]. This two-photon excitation scheme allows to independently control the axial and lateral confinement of the excitation area while taking advantage of the high depth penetration and low scattering properties [22, 23] of two-photon excitation. In temporal focusing, the spectrum of a femtosecond laser pulse is spatially dispersed by a grating. The illuminated spot on the grating is imaged via a telescope, consisting of a relay lens and the microscope objective, inside the sample (**Fig. 1A**). Thereby, the spectral components of the laser pulse overlap in time and space only in the focus region where near-diffraction limited axial confinement of the excitation is achieved leading to high two-photon excitation probability, while wide-field excitation is maintained (see Online Methods for details). Outside of the focal region the pulse is chirped which reduces the two-photon excitation probability. Temporal focusing is compatible with conventional inverted or upright microscopes (**Fig. 1A**). Using this technique a wide-field two-photon excitation area of ~ 60 × 60 µm with an axial confinement of ~ 1.9 µm is generated (Online Methods) (**Fig. 1B-C**). Furthermore, we have implemented a fast detection scheme based on an sCMOS camera coupled to a high-gain image intensifier. This eliminates the common trade-off between acquisition speed and sensitivity when comparing sCMOS and EMCCD cameras. Thereby we could retain the low noise floor of the sCMOS camera while benefiting from the signal enhancement of the intensifier (**Fig. 1D-F**).

Volumetric imaging is thus performed by scanning in the axial direction only, speeding up the volume acquisition time. For the lateral size described above and an axial range of over 30 µm, our microscope can achieve a volume acquisition rate of ~13Hz. However, due to limitations set by baseline fluorescence and signal to noise ratio of the $Ca^{2+}$-sensor, as well as tolerable expression levels in *C. elegans* (see Online Methods for details), we performed the imaging at volumetric acquisition speeds of 4-6Hz corresponding to ~5.6 Megavoxels per second, and obtained a lateral spatial resolution of 0.285 µm. At this imaging speed, the signal-to-noise ratio of the images (586 ± 111 (mean ± s.e.m.); n=10, **Fig. 1D-F**) enables us to recognize and track even dim neurons with very low basal $Ca^{2+}$- and/or expression levels (see below).

## Nuclear GCaMP captures slow and fast neuronal $Ca^{2+}$-dynamics

In order to correctly identify calcium traces from individual neurons we restricted GCaMP to the nucleus by adding nuclear localization tags [24] to G-CaMP5K [25] (**Fig. 2A**). When expressed pan-neuronally, fluorescent signals derived from nuclear-GCaMP5K (NLS-GCaMP5K) were confined to the nuclei and were absent from the neuropil. In contrast to using cytoplasmic GCaMP5K, using this approach we could identify individual cell bodies (**Fig. 2B-C**). Previous studies have reported that nuclear localized $Ca^{2+}$-indicators serve as reliable tools for detecting neuronal activity [26], but this needed confirmation in *C. elegans*. Two neuron classes called URX and BAG generate phasic $Ca^{2+}$-transients in response to rapid changes in environmental oxygen concentrations. URX is transiently activated upon oxygen upshift, while BAG is transiently activated upon oxygen downshift. In addition, BAG and URX tonically signal ambient oxygen levels [27, 28]. We tested whether NLS-GCaMP5K is a reliable reporter of both phasic and tonic signaling. We expressed NLS-CaMP5K in BAG and URX and recorded $O_2$-regulated neuronal activity via epi-fluorescence microscopy in a custom designed microfluidic device [27]. NLS-GCaMP5K fluorescence was readily detectable in the nuclei of BAG and URX (**Fig. 2D**). The changes in fluorescence reported by NLS-GCaMP5K during phasic activation of BAG and URX to $O_2$-downshift and $O_2$-upshift were similar to those reported by cytoplasmic GCaMP5K (**Fig. 2E-L**). During tonic signaling BAG and URX decrease intracellular $Ca^{2+}$ in a sustained manner when exposed to high and low $O_2$ concentrations respectively [27, 28]. NLS-GCaMP5K readily reported tonic signaling, however with reduced amplitude in $\Delta F/F_0$ when compared to cytoplasmic GCaMP5K (**Fig. 2M-R**). Our work presents NLS-GCaMP5K as a new tool that provides the ability to record fast and slow $Ca^{2+}$-dynamics within confined cell boundaries in *C. elegans*.

**Brain-wide imaging of neuronal activity**

The head of *C. elegans* contains cells in stereotypic positions and bilaterally (left-right) symmetric cell pairs represent different neuron classes. To perform brain-wide imaging in the head of *C. elegans* we first modified a microfluidic lab-on-a-chip device [27] to force curvature onto the worms, immobilizing them in a reproducible bilateral position [29], which greatly facilitates subsequent identification of neuronal classes according to their relative positions (**Fig. 3A**). Using the *Punc-31* promoter to drive NLS-GCaMP5K expression and WF-TEFO imaging we were able to record the activity patterns of 70% ± 4% (n=5) of all head-neurons (**Supplementary Fig. 1**). Expression of this construct did not affect the animals' behavior (**Supplementary Fig. 2**).

In a typical WF-TEFO imaging experiment, excitation covers the worm entirely across the dorsal-ventral axis by scanning the light disc in the axial direction (see schema in Fig. 1A). Along the anterior-posterior axis, the diameter of the excitation disc enabled us to simultaneously image the posterior part of the anterior ganglion as well as the entire dorsal, lateral and ventral ganglia (**Fig. 3B**). This encompasses a third of *C. elegans'* entire nervous system and includes about 70% of the head ganglia where most of the sensory neurons and their downstream decision making circuits are located. We performed 13 – 17 z-sections per volume with 2 µm axial step size. This was sufficient to resolve cell nuclei in the lateral as well as axial dimensions (**Fig. 3B**). High signal to noise NLS-GCaMP5K signals were recorded with 4-6 volume acquisitions per second (~80 fps). This is above the timescale of the fastest reported neuronal $Ca^{2+}$-signals in *C. elegans*, which occur on the order of seconds [27, 30], and is therefore sufficient to capture neuronal activity dynamics. We analyzed our recordings using a semi-automated image processing pipeline (Online Methods), which is able to segment a high percentage (~95%) of the neurons expressing the pan-neuronal *Punc31::NLS-GCaMP5K* construct within the imaged area (**Supplementary Fig. 1**), even at the short exposure times used in our experiments.

**Correlated and anticorrelated activity patterns in the *C. elegans* brain**

We then used this methodology to record the activity of head-neurons in non-stimulated worms. This reflects a state in which unrestrained worms would crawl and spontaneously initiate locomotion maneuvers such as turns and reversals [31]. Muscle activity was paralyzed in the animals 5 min before start of each recording, which were performed for time periods of 3-4 min. There was no measurable bleaching of NLS-GCaMP5K fluorescence during this time. In a typical recording we were able to identify and measure NLS-GCaMP5K fluorescence levels in 99 neurons across all time frames (**Fig. 3C**, **Supplementary Video 1-2**, **Supplementary Fig. 3**, **Supplementary Data 1**). To extract global features from the activity time series, we calculated the pairwise Pearson product-moment correlation coefficients between all fluorescence traces and re-ordered data traces according to agglomerative hierarchical clustering (**Fig. 3C-D**). The resulting correlation matrix (**Fig. 3D**) revealed multiple clusters composed of 2-20 correlated or anticorrelated neurons. The significance of correlations were confirmed by two bootstrapping approaches (**Supplementary Fig. 4**). We also manually confirmed that most correlations (> 90%) were reflected in the corresponding neuronal activity patterns (see Online Methods). Finally, we used a covariance analysis method to confirm the identity and correlation among highly co-active neurons (**Supplementary Fig. 5**).

Next, we focused our analysis on the neurons that showed the most prominent activities among the dataset. We empirically defined neurons as active if their $\Delta F/F_0$ signal was at any time above the mean by at least twice the standard deviation across the entire population (~36% $\Delta F/F_0$ for the example in **Fig. 3**). Manual proof checking of all individual $Ca^{2+}$-traces led us to conclude that this is a conservative cutoff. Nevertheless, half of all neurons (50 out of the 99) could be classified in this way as active, thus we conclude that at least half of the *C. elegans* brain engages in basal activity under our experimental conditions (**Fig. 4A-B** and **Supplementary Fig. 6**). We then identified the locations of some prominent neurons showing correlated activity from the entire dataset and examined their activity patterns (**Fig. 4C**). We found that some are located in bilaterally symmetric (left-right) positions, indicating that they belong to the same classes of neurons (**Fig. 4D-E**). This observation serves as a confirmation of the value of our correlation analysis and the non-random nature of the observed activities. The largest cluster contained 15-20% of all detected neurons (**Fig. 4B**), most of which are located across the lateral ganglion. The locations and morphologies of the principal neurons of this cluster are most consistent with the identities of the pre-motor interneuron classes AVA, AVE, AIB and RIB or RIM, but these identities need to be confirmed by more detailed anatomical analysis. The activity of these neurons appears anti-correlated with that of a smaller cluster (**Fig. 4B**) of neurons located in the anterior and lateral ganglia, whose positions seem to be consistent with the head motor neuron classes RMED and RMD. However, the identification of RMD was more ambiguous. Other head motor neurons that were also active during this recording seem to localize to positions most consistent with the identities of the head motor neurons SMDD or SMBD as well as SMDV. The latter two groups also showed anti-correlation with each other (**Fig. 4B-C**), an observation consistent with previous studies [32]. Our findings are reproducible across multiple animals: in the four worms that we have analyzed in detail, 2-5 periods of sustained activity across pre-motor interneurons occur within the recording times of 4 min. In all cases, these periods of sustained activity alternated with more transient neuronal activities whose locations seem to be consistent with head-motor neuron classes (**Supplementary Fig. 7-9, Supplementary Data 2-4, Fig. 5A-B, Supplementary Video 3**).

## Probing chemosensory circuits using WF-TEFO and microfluidics

Next, we set out to perform brain-wide imaging in worms under chemosensory stimulation in order to characterize the activity maps of previously characterized chemosensory circuits. We used oxygen chemosensory neurons as an example. A stimulus protocol using consecutive 30 s periods of high and low oxygen evokes robust behavioral responses (**Supplementary Fig. 2**). When this protocol was applied during an imaging session, two neuron clusters could be identified via our correlation-clustering analysis that responded with $Ca^{2+}$-transients to oxygen down- and upshifts (**Fig. 5A-B, Supplementary Video 3-4, Supplementary Fig. 9** and **Supplementary Data 2**). Their activity patterns matched those of BAG and URX (**Fig. 5A,C**). Further examination of our image data confirmed that these neurons correspond to bilaterally symmetric pairs whose morphologies and positions are consistent with the identities of URX and BAG neurons (**Fig. 5D-F**). We directly searched in the images of 11 more recordings and were able to localize BAG and URX neurons in most of them (10 and 6 out of 11, respectively). We confirmed that BAG and URX neurons reproducibly respond to $O_2$-stimuli during WF-TEFO imaging (**Supplementary Fig. 10**). Thus, when combined with microfluidic devices, brain-wide WF-TEFO imaging serves as a platform to map chemosensory neuronal circuits in small nervous systems.

## DISCUSSION

The imaging system reported here combines controlled sensory stimulation with unbiased fast volumetric neuronal recording capable of capturing---at single cell resolution---the activity of the majority of the neurons in the brain of an entire organism.

Current approaches to image neuronal populations typically read out the activity of a pre-defined fraction of cells from the total volume [10-12]. A major feature of wide-field imaging is to capture the activity of all neurons in the volume in an unbiased fashion. Hence, no a priori knowledge about the location of active neurons is required. Other variants of wide-field imaging techniques exist, most notably ones based on light-sheets [7, 14]. These have been previously applied to study *C. elegans* development [33, 34], but for that type of application the requirements on acquisition rate are less stringent than for neuronal imaging. Our technique offers several advantages: the collinear excitation configuration of our system allows for multidirectional free access to the sample when assembled on an inverted or upright microscope, thereby greatly enhancing its applicability to *in vivo* recordings in other model organisms, such as flies, fish or rodents. This configuration is also beneficial for combining imaging with electrophysiological measurements, stimulus delivery or optogenetics. Furthermore, two-photon temporal focusing excitation offers higher depth penetration, lower scattering properties [22, 23] and higher homogeneity of the axial resolution over the lateral beam profile. Two-photon imaging also avoids uncontrolled light evoked neuronal responses [35].

In our current setup, the size of the sculpted light disc was tailored to record from the head ganglia surrounding the nerve ring. By increasing the laser power and modifying the transversal shape of the beam, it would be straightforward to record from the entire anterior nervous system of *C. elegans* or brains of larger animals. The excitation area scales linearly with the average laser power, thus our current setup would allow disc sizes of up to ~200 µm in diameter, while maintaining the same power density, and thereby still avoiding tissue damage. The acquisition speed of ~80 fps in our *C. elegans* experiments is currently limited by the $Ca^{2+}$-reporter brightness and not by the available laser power, the speed of the axial scanning or the camera frame rate. Our current instrumentation allows imaging speeds of up to 200 fps. Therefore, our approach can potentially be extended to volumetric imaging of larger brains and recording of spiking neurons that exhibit faster $Ca^{2+}$-dynamics.

A crucial aspect of our study was the development of NLS-GCaMP5K as an imaging probe for recording the activity from dense neuronal tissues in *C. elegans*. Cytoplasmic $Ca^{2+}$-signals in *C. elegans* are an appropriate proxy for assessing neuronal activity (**Supplementary Table 1**). Electrophysiological studies have revealed that *C. elegans* neurons exhibit $K^{+}$ and $Ca^{2+}$ based graded and regenerative potentials [36]. Synaptic release can also occur in a graded fashion [37]. Based on these findings a rise in intracellular $Ca^{2+}$ can be assumed to coincide with membrane depolarization, which is detected in a graded fashion by downstream chemical and electrical synapses. $Ca^{2+}$-signals in *C. elegans* occur at time scales of up to 1 Hz [27, 30]. Given these slow dynamics, we hypothesize that nuclear-localized G-CaMP5K should not be limited by possible longer diffusion rates of $Ca^{2+}$ to the nucleus. Indeed, here we report that nuclear $Ca^{2+}$-signals in *C. elegans* are very similar in kinetics and amplitude to their cytoplasmic counterparts, at least in the two sensory neurons that we analyzed. Our findings might not apply to all *C. elegans* neurons. Furthermore, our pan-neuronal imaging results indicate that NLS-GCaMP5K reports activity from at least 50% of all neurons, which is a very conservative estimate as many neurons, such as other chemosensors in the head, are expected to be inactive in our experimental conditions. Some *C. elegans* neurons signal via $Ca^{2+}$-domains at the level of their processes [32, 38]. A nuclear $Ca^{2+}$-reporter is obviously unable to capture these dynamics, which occur in the nerve ring neuropil. Therefore, the approach presented here is aimed at assessing the instantaneous activity of neurons in comparison to the rest of the brain, and should not be understood as an approach aimed at replacing detailed single-cell studies.

An intriguing feature that we observe is correlated periods of sustained activity among a large fraction of the recorded neurons (23 neurons), the locations of some of which are consistent with interneuron identities implicated in behavioral decisions, such as switching between forward and backward locomotion [31, 39]. The dynamic patterns of $Ca^{2+}$ signals that we observed in these cells are similar to those that were recorded in non-paralyzed worms [39-

[41]. Despite the irregularity of activity patterns that we observed, we found a high level of correlated activity, something that may not be too surprising given the high level of interconnectivity (via chemical or electrical synapses) revealed by the worm's wiring diagram [9]. Additionally, our results confirm a previously described antagonism between head motor neurons that innervate opposite parts of the musculature [32]. In conclusion, our approach enables extracting characteristic clusters of identifiable neurons that are functionally and anatomically linked. Taken together, one may speculate that system wide correlations rather than isolated activities of individual neurons underlie changes in locomotion state. We would like to point out, however, that our studies used a cholinergic agonist for paralysis, a condition that might lead to aberrant network states [42]. Questions relating to whether widespread correlations occur in non-paralyzed worms, and what the functional relevance of the observed correlations and anti-correlations are, as well as the unambiguous identification of all neurons in the correlation-clusters, will be explored in future studies.

## Acknowledgements

We thank J. Akerboom and L. Looger (Janelia Farm Research Campus, Howard Hughes Medical Institute) for valuable information on characteristics of GCaMP variants and for constructs, M. Colombini for manufacturing of mechanical components, E. Sanchez, B. Bathellier, G. Haunert, D. Aschauer for helpful discussions and valuable input, M. Palfreyman, H. Kaplan, S. Kato and C. Bargmann for critically reading the manuscript, R. Latham, M. Sonntag, D. Grzadziela and S. Skora for technical support. R.P. acknowledges the VIPS Program of the Austrian Federal Ministry of Science and Research and the City of Vienna as well as the European Commission (Marie Curie, FP7-PEOPLE-2011-IIF). The research leading to these results has received funding from the European Community's Seventh Framework Programme (FP7/2007-2013) / ERC grant agreement n°.281869 - *elegansNeurocircuits*, the Vienna Science and Technology Fund (WWTF) project VRG10-11, the Human Frontiers Science Program Project RGP0041/2012, the Research Platform Quantum Phenomena and Nanosclae Biological Systems (QuNaBioS) and the Research Institute of Molecular Pathology (IMP). The IMP is funded by Boehringer Ingelheim.

## Author Contributions

T.S. and R.P. designed and performed experiments and analyzed data, R.P. and A.V. designed and built the imaging system, T.S. and M.Z. designed and characterized NLS-GCaMP5K, and designed and validated the microfluidic device, K.A. wrote analysis software and analyzed data. M.Z. and A.V. designed experiments, conceived of and led the project. T.S., R.P., M.Z. and A.V. wrote the manuscript.

## Competing Financial Interests

The authors declare no competing financial interests.

## Figure legends

**Figure 1. Volumetric fluorescence imaging using wide-field two-photon light sculpting.**
**(A)** Schematic depicting of the light-sculpting microscope and microfluidic sample holder. The pulses at the bottom sketch the geometric dispersion in temporal focusing as a function of axial position. The inset on the top right is an artistic rendering of a *C. elegans* head, indicating axially scanned light discs and the imaged region. Neurons URX and BAG are also depicted. Scale bar is 15 µm. DC, dichroic mirror; OPA, optical parametric amplifier; sCMOS, scientific complementary metal–oxide–semiconductor **(B)** Lateral characteristics of the excitation profile measured on a homogenous fluorescent test sample showing Gaussian intensity pixel distribution. **(C)** Optical sectioning capability of the light disc generated by temporal focusing. Measured (solid dots) and fitted (line) intensity of an axially displaced, 100 nm-thin fluorescent test sample. **(D-E)** Maximum-intensity projection (MIP) of typical volumetric recordings in *C. elegans* with **(D)** and without **(E)** a high-gain image intensifier in front of the sCMOS camera. The red boxes indicate regions of interest (ROI) used in determining the signal-to-noise. **(F)** SNR as a function of intensifier gain (AU) using an identical worm sample and exposure time as in (D), (E). The same brightest neurons (n = 10) were used across recordings for the analysis. A – anterior; P – posterior; D – dorsal; V – ventral; OPA – optical parametric amplifier.

**Figure 2. *in-vivo* characterization of NLS-GCaMP5K. (A)** Schematic of NLS-GCaMP5K. **(B-C)** Maximum intensity projections of fluorescence-image-stacks acquired with a spinning disk confocal microscope. Shown are head ganglia pan-neuronally expressing cytoplasmic G-CaMP5K **(B)** versus nuclear G-CaMP5K **(C)**. Arrows indicate the locations of the nerve ring neuropil. **(D-R)** $Ca^{2+}$-imaging of BAG and URX neurons by epi-fluorescence microscopy. **(D-F)** NLS-GCaMP5K fluorescence in stimulated chemosensory neurons. Colorbar on the right indicates scaled fluorescence intensities. **(D)** Baseline (21% $O_2$) fluorescence levels in $O_2$-sensing neurons BAG and URX. **(E)** $O_2$-downshift evoked response in BAG. **(F)** $O_2$-upshift evoked response in URX. **(G-R)** Averaged $Ca^{2+}$-transients in BAG and URX and quantifications. Traces indicate mean $\Delta F/F_0$. Shading indicates s.e.m. White and blue backgrounds indicate periods at 21% $O_2$ and 4% $O_2$, respectively. The number of worms analyzed in each case is noted in the graphs. **(G-I)** $O_2$ downshift evoked $Ca^{2+}$-transients in BAG expressing cytoplasmic **(G)** or nuclear **(H)** G-GAMP5K and quantification of mean peak responses **(I)**. **(J-L)** $O_2$ upshift evoked $Ca^{2+}$-transients in URX expressing cytoplasmic **(J)** or nuclear **(K)** G-GAMP5K and quantification of mean peak responses **(L)**. **(M-O)** $O_2$-regulated tonic $Ca^{2+}$-signaling in BAG measured with cytoplasmic **(M)** or nuclear G-CaMP5K **(N)** and quantification of mean response **(O)**. **(P-R)** $O_2$-regulated tonic $Ca^{2+}$-signaling in URX measured with cytoplasmic **(P)** or nuclear **(Q)** G-CaMP5K and quantification of mean response **(R)**. Bar graphs in the right panels show mean $\Delta F/F_0$ at time intervals indicated by red bars below the traces in the left and middle panels. Asterisks indicate significance levels by t-test (*p=0.0147, ***p = 0.0006, ns not significant). All scale bars represent 10 µm.

**Figure 3. Brain-wide WF-TEFO $Ca^{2+}$-imaging in *C. elegans*. (A)** Schematics of the microfluidic PDMS device. **(I-II)** Red and blue sketches indicate worm immobilization and gas delivery channel respectively. **(II)** Cross-section (not to scale). **(III)** Phase contrast image of immobilized worm inside the device. The head (tail) is at the bottom (top). Ventral is shown by the vulva's location (blue arrow). **(B)** WF-TEFO imaging of head region of a *Punc-31::NLS-GCaMP* worm. **(I)** Maximum intensity projection of 14 z-planes. Dashed lines outline the locations of head ganglia as shown in (II). Scale bar represents 10 µm and refers to panels I,III-V. **(II)** Schematic of the left anterior head ganglia modified from ref. [9]. Black lines outline neuronal nuclei. Grey lines outline the anterior ganglia as indicated. Green area indicates pharynx. **(III)** Single z-plane (z = 2 µm). Dashed lines indicate *y-z* and *x-z* cross-sections shown in **(IV)** and **(V)** respectively. Arrows in (III-V) indicate example neurons each seen in two projections. The crosses indicate worm orientation (A – anterior, P – posterior, D – dorsal, V – ventral, L – left, R - right). **(C)** Activity of 99 neurons from the same worm as in B, imaged volumetrically at 5 Hz for 200s. Each row shows a heat-plot of NLS-GCaMP5K fluorescence time series. Color indicates percent fluorescence changes *($\Delta F/F_0$)*. Colorbar on the left indicates scaling. X-axis represents elapsed recording time. **(D)** Matrix showing correlation coefficient (*R*) calculated from all time-series shown in (C). Color indicates the degree of correlation. Colorbar on the right indicates

scaling. The data in C-D are grouped by agglomerative hierarchical clustering. Corresponding rows in (C) and (D) are aligned. **Supplementary Fig. 3** provides high resolution images of panels (C-D) that indicate the neuron ID numbers consistent with **Supplementary Image 1** and **Supplementary Data 1**.

**Figure 4. Time-series correlations between neurons (A)** NLS-GCaMP5K fluorescence time series of active neurons from Figure 3C. Color indicates percent fluorescence changes *($\Delta F/F_0$)*. Colorbar indicates scaling. X-axis represents elapsed recording time. **(B)** Matrix of correlation coefficients (*R*). Color indicates the degree of correlation. Left colorbar indicates scaling. Data in (A-B) are grouped by agglomerative hierarchical clustering. Corresponding rows in (A) and (B) are aligned. **Supplementary Fig. 6** provides high resolution images of panels (A-B). **(C)** Selected traces of neurons. Numbers (Neuron ID) in this and all subsequent panels identify neurons consistent with **Supplementary Fig. 3** and **6**, **Supplementary Image 1** and **Supplementary Data 1**. Colors correspond to the clusters indicated by colored bars in (A-B). Dashed lines indicate time points corresponding to images in (D). **(D)** Examples for *y-z* (I-III) and *x-z* (IV) projections of some neurons shown in (C). Crossed arrows show worm orientation (A – anterior, P – posterior, D – dorsal, V – ventral, L – left, R - right). Perpendicularly aligned arrows indicate projection orientations. **(I-III)** *y-z* projections through pairs of correlated bilateral symmetrically oriented neurons. The correlation coefficients are $R_{5/87}$=0.986 (I), $R_{12/96}$=0.893 (II) and $R_{92/43}$=0.8414(c). **(III)** *x-z* projection across the boundary between the lateral and ventral ganglia showing a group of 6 synchronously active neurons. Colorbars show fluorescence intensities. **(E)** Schematic of the left head ganglia. Neuron colors match with the colors of traces in (C) and bars in (A-B). Dashed lines outline the focal plane of each projection in (D) and are labelled by the corresponding panel number. All scale bars represent 5 µm.

**Figure 5. WF-TEFO $Ca^{2+}$-imaging in worms during chemosensory stimulation. (A)** Activity of 69 neurons under consecutively changing $O_2$-concentrations. Rows are heat-plots of NLS-GCaMP5K fluorescence time series. Color indicates percent fluorescence changes (*$\Delta F/F_0$*). Colorbar indicates scaling. X-axis represents elapsed recording time. $O_2$-concentrations shifted as indicated by dashed lines. **(B)** Matrix showing correlation coefficient (*R*) calculated from data in (A). Color indicates the degree of correlation. Colorbar indicates scaling. The data in A-B are grouped by agglomerative hierarchical clustering. Corresponding rows in A and B are aligned. **Supplementary Fig. 9** provides high resolution images of panels (A-B) that indicate the neuron ID numbers consistent with (C-F), **Supplementary Fig. 9, Supplementary Image 2** and **Supplementary Data 2**. **(C)** Fluorescence traces of oxygen sensory neurons. Colors correspond to the neurons labeled by colored bars in (A-B). $O_2$-concentration shifts are indicated on top and by dashed lines. **(D)** Maximum intensity *x-y* projection of images at time point t = 45s. Dashed lines indicate cut sections for the projections in (E-

F) as indicated. Crossed arrows indicate worm orientation. Numbers are neuron IDs. **(E-F)** Examples for *x-z* (left panels) and *y-z* (right panels) projections of BAG **(E)** and URX **(F)** neurons at indicated time points. Numbers are neuron IDs. The crossed arrows show the worm orientation (A – anterior, P – posterior, D – dorsal, V – ventral, L – left, R - right). Perpendicularly aligned arrows indicate the projection orientation. Colorbars show fluorescence intensities. **.**All scale bars represent 10 µm.

## ONLINE METHODS

### Fast two-photon volumetric imaging using temporal focusing.

In temporal focusing, the frequency spectrum of an ultrafast (femto-second) laser pulse is spatially dispersed by a grating and the spot on the grating is imaged inside the sample via a relay lens and the objective (**Fig. 1A**). The spectral components of the laser pulse will only spatially and temporally overlap in the focus region of the objective where the original pulse duration will be restored. Outside the focal region the pulses are spatially dispersed and temporally chirped. Since the probability of exciting a fluorescent molecule scales inversely with laser pulse duration, this leads to a confinement of the excitation in a relatively small axial region, while wide-field excitation can be achieved in the lateral dimensions. However, as the excitation light in temporal focusing is distributed over a wider area than in conventional two-photon scanning microscopy, higher peak pulse energies are required. Thus, we use a low-repetition rate regenerative amplifier (Legend Elite, Coherent) seeded by a broadband Ti:Sapph oscillator (Mantis, Coherent, 75MHz, 20fs). The output of the amplifier (0.5 mJ, 10 kHz, 35 fs) is further tuned to the optimal excitation wavelength of GCaMP5K (970 nm) by employing an Optical Parametric Amplifier (TOPAS-C, Light Conversion). Its output (typically 20 µJ, 120 fs) is attenuated by a combination of a motorized half-wave plate and polarizing beam-splitter to yield ~2 µJ at the sample. The beam diameter is expanded to ~10 mm diameter with a telescope before the grating (830 lines/mm, Newport). The relay lens (f = 600 mm) and objective (Olympus UPLAN-APO 40x 1.3NA) form an excitation disc of ~60 µm ($1/e^2$) diameter in the focus of the objective. We typically use an average power of 15-20 mW at the sample, which is comparable to power densities of conventional two-photon scanning microscopes, which in our experience did not induce photodamage in the worm, even over 4 min of continuous imaging. Photodamage appeared as changes in tissue morphology that is seen seen in bright field images when laser power approached 40mW. Scanning in the axial direction is performed using a high-speed piezo stage (Nano-OP65HS and Nano-Drive 85, Mad City Labs). The individual (2 µm) steps of the piezo-element in between camera exposures are driven sinusoidally and image acquisition is triggered after each individual z-positions is reached. The overall movement of the piezo over multiple z-stacks resembles a saw-tooth curve (in contrast to triangular movement). The piezo is synchronized to the camera via a custom LabView script. On the detection side, a custom dichroic mirror followed by multiphoton filter (HC680/SP-25, Semrock) separate the fluorescence signal from the excitation light. The signal is further split into two channels using a dichroic mirror (HC BS 560, Semrock) and bandpass filters (BrightLine 525/50, for GCaMP and BrightLine 605/70, for red fluorophore like DsRed2) inside a wavelength separator (Optosplit, Cairn Research). Fluorescence is recorded in a wide-field manner by employing a high-gain, gated GenII Image Intensifier (C9016-21, Hamamatsu, effective pixel size 10$\mu$m, 17mm diameter) whose output is imaged via a 1:1 relay lens onto the sCMOS camera chip (Neo, Andor, 6.5$\mu$m pixel size). This combination maintained high-signal-to-noise even at short exposure times of 10ms (**Fig. 1D-F**). The intensifier creates spurious noise apparent as bright pixels at high gain in single frames. These were easily removed during image post processing by pixel-wise comparison of two adjacent frames in time and correcting to the minimum value of each pixel pair. Given the effective magnification (44.4x) of our detection optics, the read-out area of 512x512 pixels on our sCMOS camera chip translates to a field of view of 75x75$\mu$m at the sample. Given the sustained read-out speed of the sCMOS camera in this configuration (2.56 ms per image) and the finite translation time of the piezo stage (~2.5 ms per move), we are thus able to record volumetric image slices at a rate of ~200 Hz, corresponding to 22 Megavoxels per second. In the actual experiments with *C. elegans*, because of longer exposure times and the fact that the excitation light disc does not cover the whole read-out area, this decreases to effectively 5.6 Megavoxels, or ~4-6 volumes per second, depending on the thickness of the worm. Image acquisition is performed with custom-written LabView and Andor Solis Basic scripts. Synchronization between the piezo stage and camera is done via TTL triggering. The sCMOS camera was operated at the following settings: Kinetic series, external shutter, 10 ms exposure time, 16bit, 560M-F read-out speed, rolling shutter, no binning, (512 x 1024) or (512 x 512) ROI, spooling to SSD hard-disc.

**Image processing and data analysis.**

Image post-processing on an average of 16,000 images (32GB) per experiment were done using custom-written scripts (MetaMorph, MolecularDevices, *Universal Imaging*). The pixel intensities of images are first corrected by dividing them with normalized reference excitation images that were taken at each day of data acquisition. Figure 1B shows an example of such an image. This correction step accounts for the spatial inhomogeneity of the $TEM_{00}$ mode of the excitation beam.

To extract the mean fluorescence intensity of individual neurons from the raw images at each time-point, we used the multi-dimensional data analysis tool in MetaMorph software (*Universal Imaging*) to first manually assign square regions to all detectable neurons in every single plane (**Supplementary Images 1-4**). The size of these regions was between 9x9 and 12x12 pixels and they were located within the borders of the visible nucleus. For each z-plane we considered all time-points in order to also detect neurons, which due to low baseline fluorescence and/or large changes in fluorescent intensity, were not clearly visible over the whole time-series. For each individual neuron we chose the z-plane in which fluorescence intensity was the highest for tracking in 2D. Whenever we observed slight movement in z of one neuron, we tracked and measured its intensity of on the maximum intensity projection of both planes across which the neuron was moving. Time-series data were measured semi-automatically from the image time-stacks using a customized tracking script based on the MetaMorph track objects function (template match algorithm). This script tracks one selected object per plane (i.e. one bright nucleus which is clearly visible over all time-points) and adjusts the x-y positions of up to eight regions simultaneously according to the movement of the tracked object. The script logs the average fluorescent intensity of each single region for every time-point. Subsequent data and analysis was carried out using custom scripts written in MatLab (MathWorks) and Mathematica (Wolfram Research). To correct for any remaining artifacts caused by movement drift across the Gaussian excitation pattern we introduced another correction step. This was necessary because the excitation pattern is not measurable in the background, thus simple normalization to background was not possible. We therefore manually selected inactive neurons as references and assigned other close by (within a radius of 30 $\mu$m) neurons to them. Each fluorescence trace $F$(t) was normalized to its reference neuron by division of the fluorescence traces and multiplication by a scaling factor, which was determined as the ratio between the mean fluorescences within each pair: $F(\mathrm{t})_{\mathrm{correct}} = F(\mathrm{t})/F(\mathrm{t})_{\mathrm{reference}}$ * $(\mathrm{mean}(F(\mathrm{t}))/\mathrm{mean}(F(\mathrm{t})_{\mathrm{reference}}))$. To extract $\Delta F/F_0$, we calculated $\Delta F/F_0 = 100 * (F(\mathrm{t})_{\mathrm{correct}} - F_0)/F_0$ with $F_0$ being the mean fluorescence intensity of each corrected trace. We chose the mean because this allowed for the best comparison across whole datasets, in which baseline fluorescence intensities varied substantially.

Correlation coefficients were calculated in MatLab. Manually proof checking confirmed >90% of all correlations; the remaining was due to imaging artifacts such as emission light scatter from adjacent neurons and movement. However, an unbiased correlation analysis is very sensitive and can additionally detect also weak but correlated changes in fluorescence when the sample slightly moves across the excitation beam pattern. We therefore also used MatLab to calculate the covariance of the signals; this approach emphasizes relationships of highly co-active neurons. This method generally confirmed the highest correlations in the correlation matrix (**Supplementary Fig. 5**). Agglomerative hierarchical clustering was performed on the correlation and covariance matrices using the MatLab linkage function based on a Euclidean distance measure.

## Nuclear expression of GCaMP5K in *C. elegans*.

To restrict expression of GCaMP5K to the nucleus two nuclear localization signals (NLS) were added as reported previously for GFP [24]. One NLS of simian virus 40 (SV40) was fused to the N terminus and a putative NLS of the *C.elegans* transcription factor EGL-13 to the C terminus. GCaMP5K was amplified with primers 5'-AATATTGCTAGCATGGCTCCAAAGAAGAAGCGTAAGG TAATGGGTTCTCATCATCAT-3' (including a *NheI* site and $NLS^{SV40}$) and 5'-ATTATTGGCGCCCTTCGCTGTCATCATTTG-3' (introducing a *NarI* site right before the stop codon). The 75-bp NLS of *egl-13* was amplified with primers 5′-ATGGC GCCATGAGCCGTAGACGAAAAGCG-3′ (including *NarI* site) and 5′-ATGGTACCTTA ATTTTCAACTTCCTTGGCAAGC-3′ (including KpnI site). $NLS^{SV40}$-GCaMP5K and $NLS^{EGL-13}$ DNA fragments were ligated together into *NheI/KpnI* cut pSM vector to yield plasmid pTS31 (pSM/NLS-GCaMP5K). For generating a pan-neuronal expression construct, 6.8 kb upstream genomic region of the *unc-31* gene was then subcloned to yield plasmid pTS36 (pSM/Punc-31::NLS-GCaMP5K). The construct was injected at 30 ng/µl together with a coelomocyte co-injection marker into wild type *C. elegans* Bristol strain N2 to obtain strain ZIM294 (genotype: *mzmEx199[Punc-31::NLSGCaMP5K; Punc-122::gfp]*). Animals were maintained under standard conditions.

### ***In vivo* calcium imaging of *C. elegans* head ganglia neurons.**

Microfluidic two-layer PDMS devices (**Fig. 3A**) were constructed as previously described [27]. $O_2$ delivery was accomplished by attaching the gas inlet of the imaging device via a T-connector to two three-way valves controlling the intake of two tanks containing pressurized gas pre-mixtures of $O_2$ and nitrogen. The two valves were automatically controlled by the ValveBank 8II (Automate Scientific, Inc.) as described before [27] Both gas mixtures were constantly flowing but only one mixture was led into the flow chamber at any time. The gas flow rate was adjusted to yield a pressure of 0.5 psi at the outlet of the flow chamber. The worm channel was connected to a reservoir containing S-Basal buffer with 5 mM tetramisole, an acetylcholine receptor specific agonist that was used to mildly paralyze the animal's muscles to reduce motion. All components were connected with TYGON tubing (0.02 in ID, 0.06 in OD; Norton) or polyethylene tubing (0.066 in ID, 0.095 in OD; Intramedic) using 23G Luer-stub adapters (Intramedic).

For $Ca^{2+}$ imaging, adult worms (one to four egg stage) expressing NLS-GCaMP5K were loaded into the worm channel. Animals were transferred into a drop of S-Basal containing 5 mM tetramisole on a food free nematode growth medium (NGM) plate. By applying a short vacuum to the worm outlet, animals were sucked up into TYGON tubing, which was afterwards connected again to the worm inlet to position the worm in the curved channel. Imaging data in Figure 2D-R were acquired with an epi-fluorescence microscope at 10 fps. In our WF-TEFO imaging experiments, two critical parameters had to be carefully chosen. First, we operated at laser powers approximately one half of the level that induces photo-damage in the worm's tissue (15-20 mW vs ~40 mW average input power). Second, GCaMP5K signal-strength is limited by two factors: the low baseline fluorescence of GCaMP makes neurons with low $Ca^{2+}$-levels difficult to detect. Further, high expression levels of GCaMP can lead to $Ca^{2+}$-buffering, which affects cell function [43]. Interestingly, we experienced that NLS-GCaMP5K is less deleterious in this respect when compared to cytoplasmic GCaMP versions. The following observations led us to this conclusion: attempts to obtain transgenes expressing cytoplasmic GCaMP5K in a pan-neuronal fashion led to very low yield of stable expressers as from the F1 generation, i.e. GCaMP5K expression in transformed animals in the F2 generation was usually very dim and occurred in a mosaic fashion. Only a few transgenic lines were suitable for further studies; one of them (ZIM423, genotype: *mzmEx285[Punc-31::GCaMP5K; Punc-31::tdTomato-His-58]*) was used for generating the data in Figure 2B. These observations are consistent with previous experience in generating GCaMP5K transgenes using other cell specific promoters. Taken together, our experience is that cytoplasmic GCaMP5K might be toxic when expressed at too high levels. When NLS-GCaMP5K plasmids were injected at the same concentrations, the yield of transgenes was much higher, expression levels as assessed by fluorescence intensities were higher and less mosaic expression was observed.

## Confocal imaging

Confocal images (**Fig. 2B-C**) were acquired with a Perkin Elmer spinning disk confocal microscope. Exposure time was set to 20-60 ms per plane. About 100 z-planes were acquired with to provide high-resolution images.

## Methods-only References

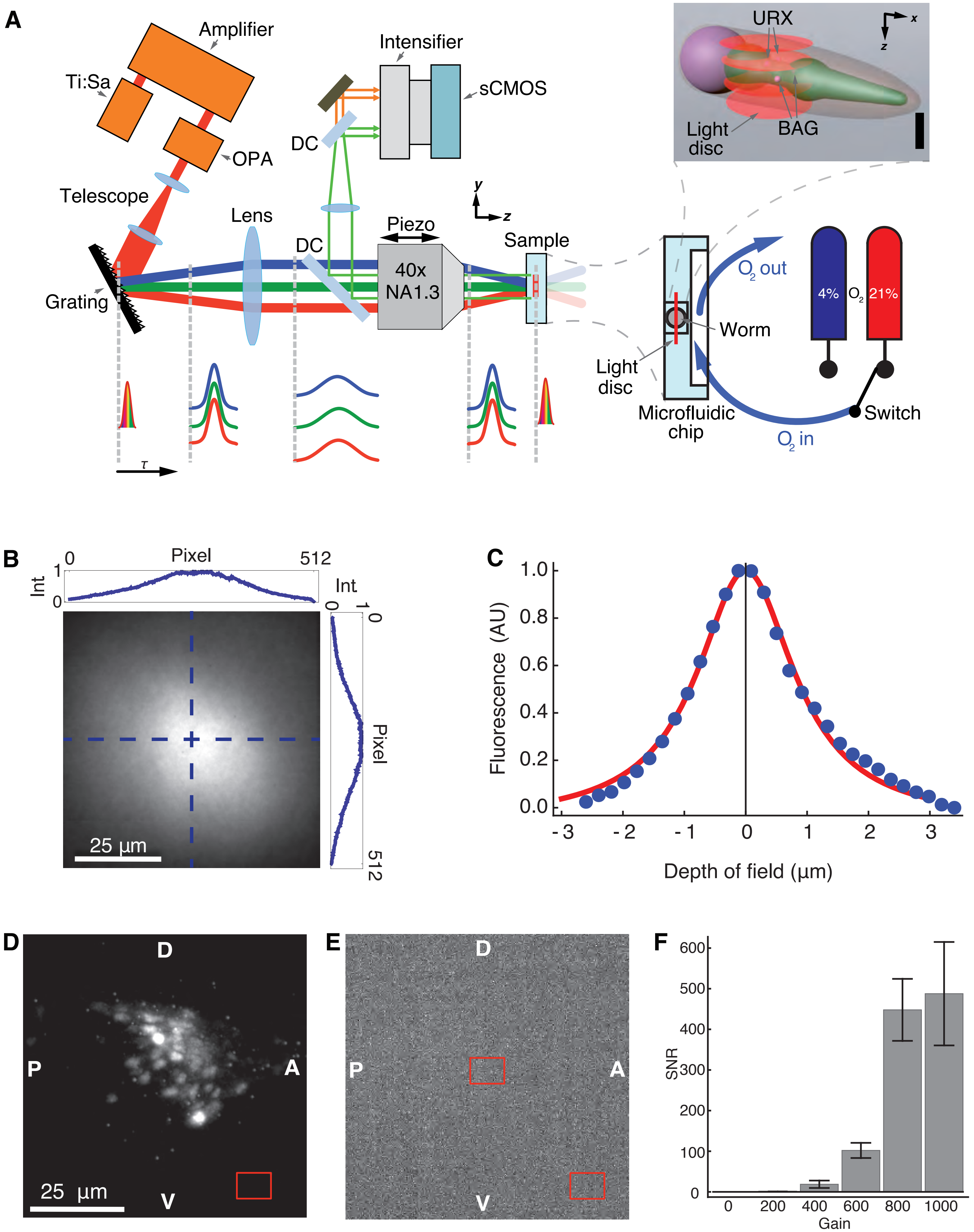

Figure 2

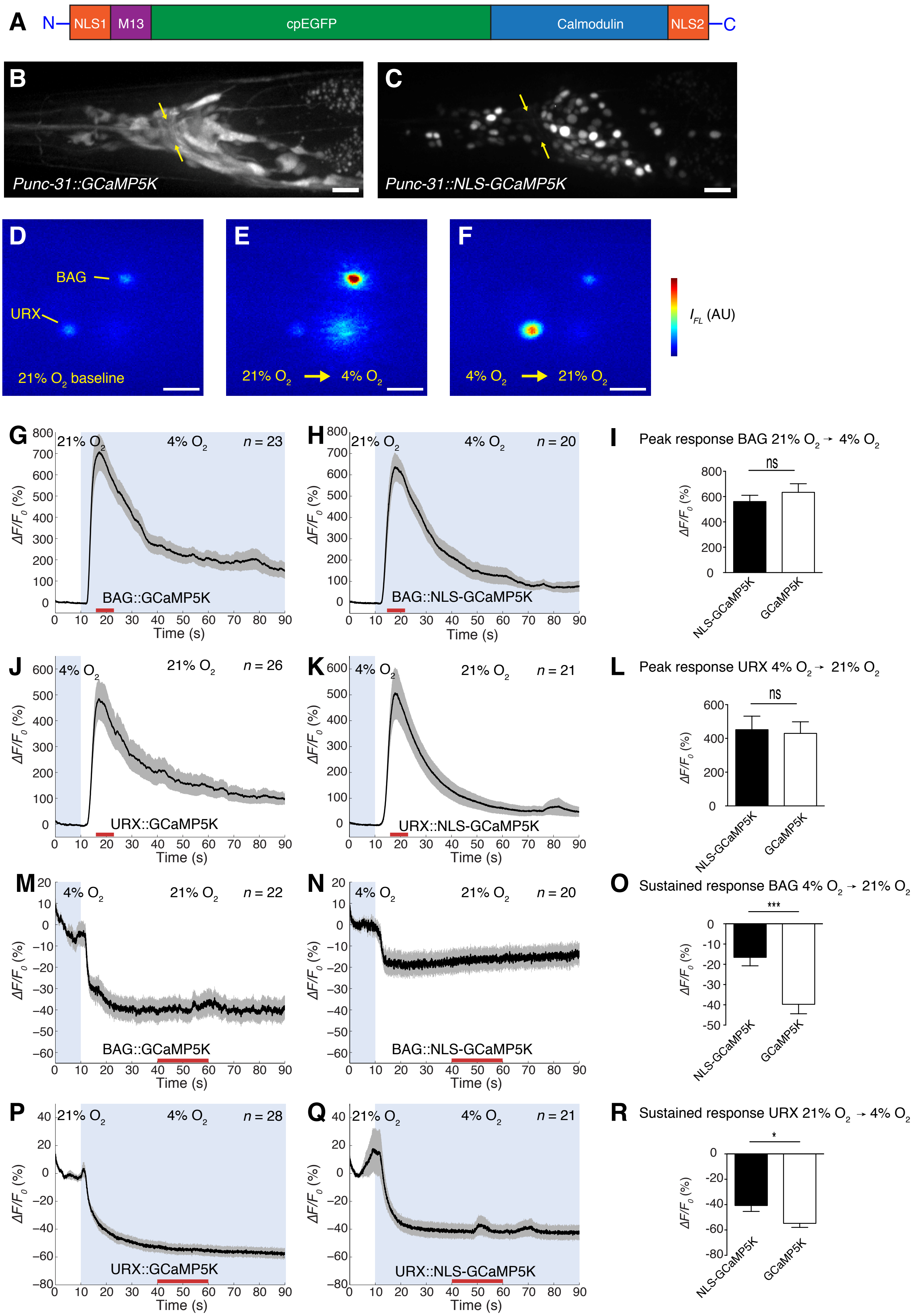

A
N
NLS1
M13
cpEGFP
Calmodulin
NLS2
C
B
Punc-31::GCaMP5K
C
Punc-31::NLS-GCaMP5K
D
BAG
URX
21% O2 baseline
E
21% O2 → 4% O2
F
4% O2 → 21% O2
IFL (AU)
G
21% O2
4% O2
n = 23
BAG::GCaMP5K
H
21% O2
4% O2
n = 20
BAG::NLS-GCaMP5K
I
Peak response BAG 21% O2 → 4% O2
ns
NLS-GCaMP5K
GCaMP5K
J
4% O2
21% O2
n = 26
URX::GCaMP5K
K
4% O2
21% O2
n = 21
URX::NLS-GCaMP5K
L
Peak response URX 4% O2 → 21% O2
ns
M
4% O2
21% O2
n = 22
BAG::GCaMP5K
N
4% O2
21% O2
n = 20
BAG::NLS-GCaMP5K
O
Sustained response BAG 4% O2 → 21% O2
***
P
21% O2
4% O2
n = 28
URX::GCaMP5K
Q
21% O2
4% O2
n = 21
URX::NLS-GCaMP5K
R
Sustained response URX 21% O2 → 4% O2
*
ΔF/F0 (%)
Time (s)

Figure 3

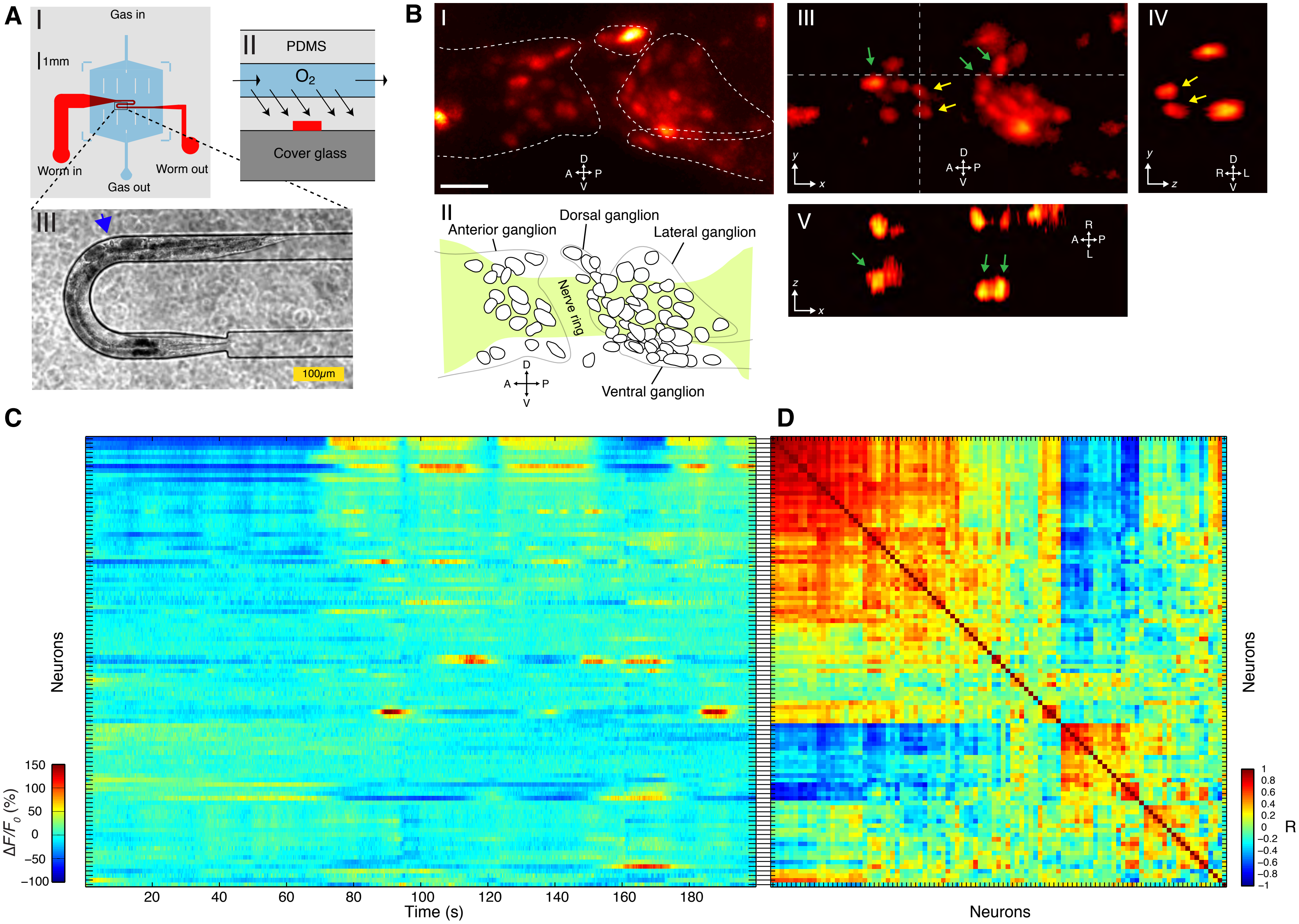

Figure 4

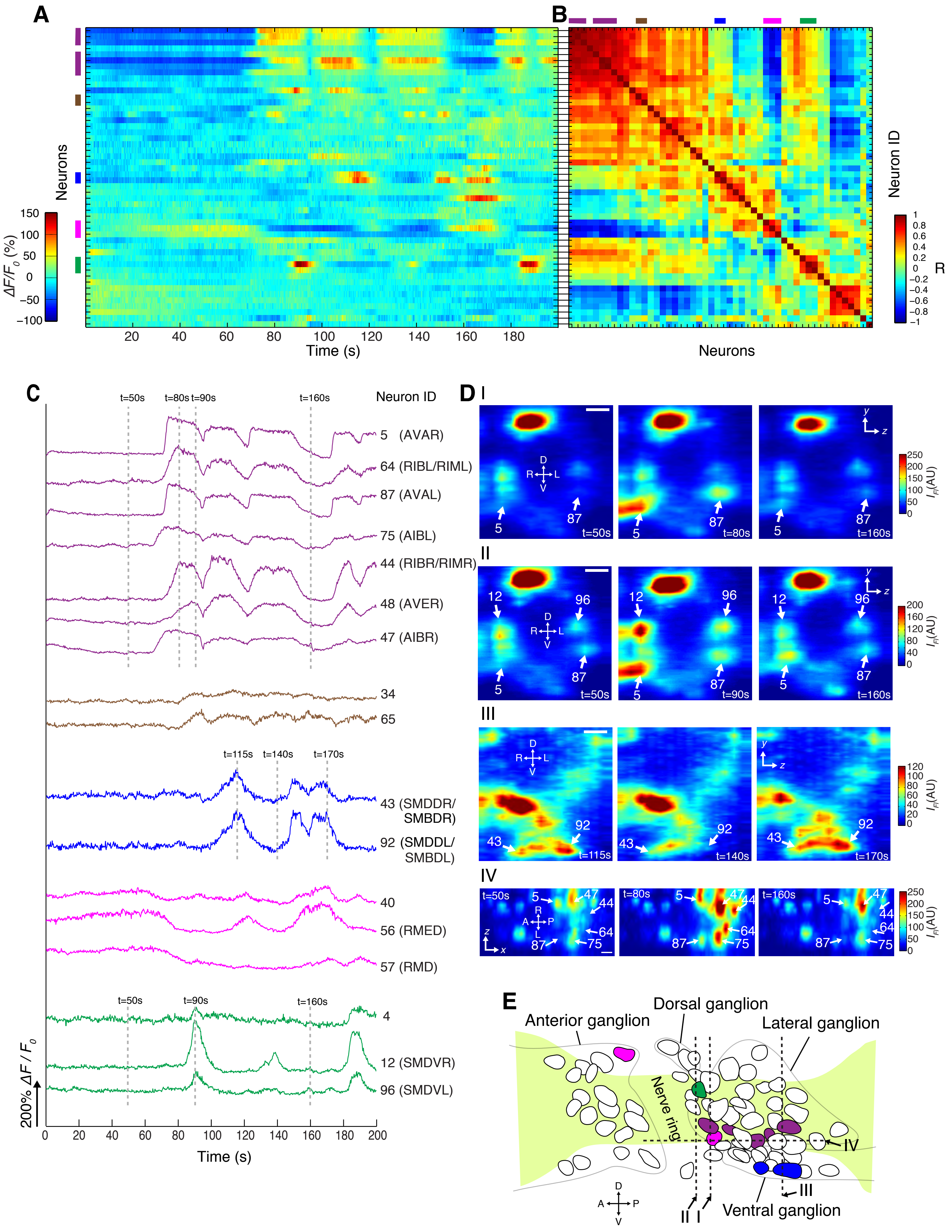

Figure 5

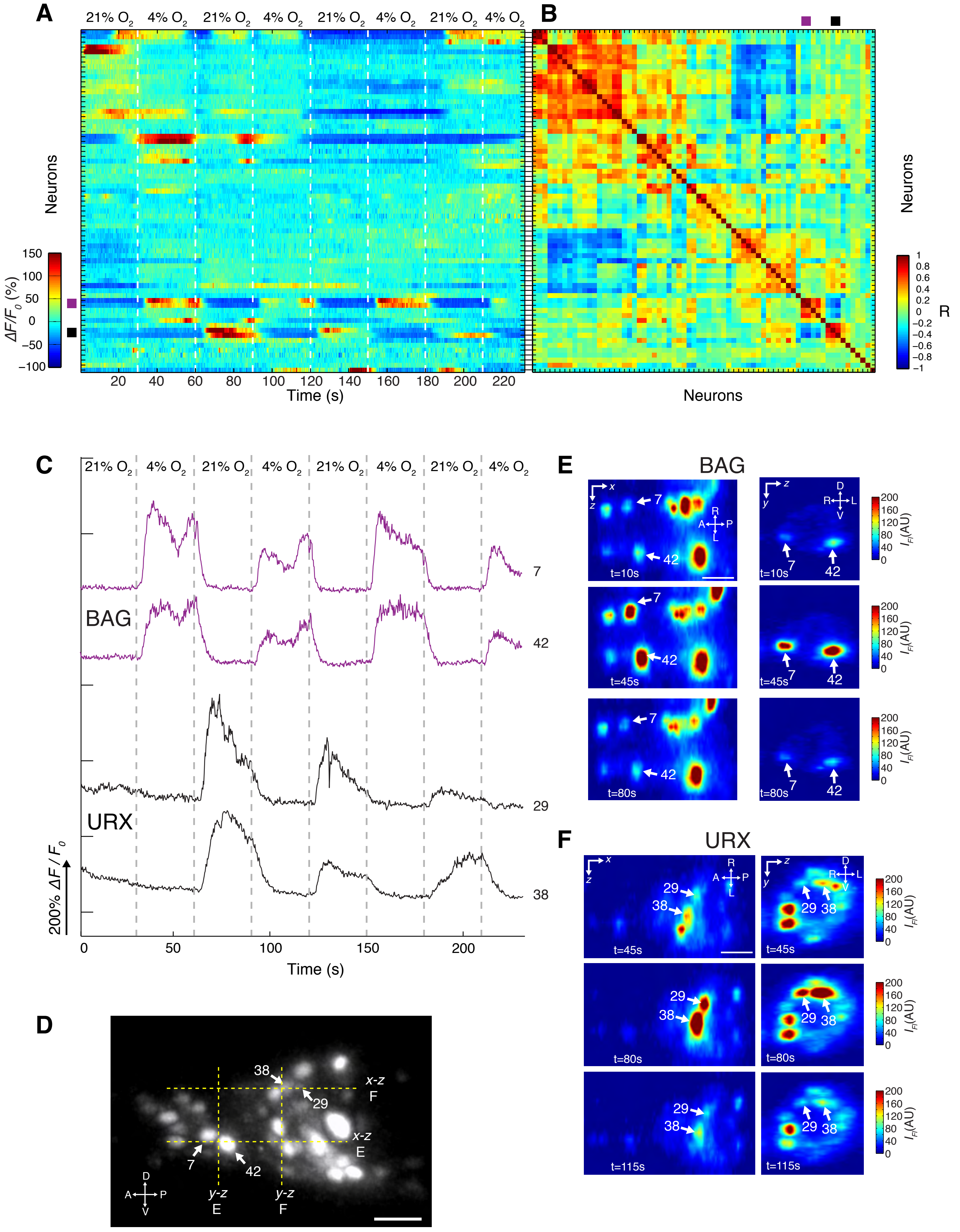
A
21% O2 4% O2 21% O2 4% O2 21% O2 4% O2 21% O2 4% O2
Neurons
ΔF/F0 (%)
150
100
50
0
−50
−100
20 40 60 80 100 120 140 160 180 200 220
Time (s)
B
Neurons
R
1
0.8
0.6
0.4
0.2
0
−0.2
−0.4
−0.6
−0.8
−1
Neurons
C
21% O2 4% O2 21% O2 4% O2 21% O2 4% O2 21% O2 4% O2
BAG
7
42
URX
29
38
200% ΔF / F0
0 50 100 150 200
Time (s)
D
38
29
x-z
F
x-z
E
7
42
y-z
E
y-z
F
D
A P
V
E
BAG
t=10s
t=45s
t=80s
IFl(AU)
F
URX
t=45s
t=80s
t=115s

# Supplementary Information

## Brain-wide 3D imaging of neuronal activity in *Caenorhabditis elegans* with sculpted light

Tina Schrödel[1,5], Robert Prevedel[1,2,3,5], Karin Aumayr[1,4], Manuel Zimmer[1] & Alipasha Vaziri[1,2,3]

[1] Research Institute of Molecular Pathology, Vienna, Austria
[2] Max F. Perutz Laboratories GmbH, University of Vienna, Vienna, Austria
[3] Research Platform Quantum Phenomena & Nanoscale Biological Systems, Vienna, Austria
[4] Institute of Molecular Biotechnology, Austrian Academy of Sciences, Vienna, Austria

[5] These authors contributed equally to this work.

Correspondence should be addressed to M.Z. (zimmer@imp.ac.at) or A.V. (vaziri@imp.ac.at).

## Supplementary Figures

**Supplementary Figure 1.** Detectable expression of neurons with *Punc31-* pan-neuronal promoter constructs.

**Supplementary Figure 2.** Behavioral responses to repetitive $O_2$-shifts of wild type and *Punc-31::NLS-GCaMP5K* worms.

**Supplementary Figure 3.** High resolution image of Figure 3C-D indicating neuron ID numbers.

**Supplementary Figure 4.** Bootstrapping approaches confirming correlation coefficients.

**Supplementary Figure 5.** Clustering of data according to covariance.

**Supplementary Figure 6.** High resolution image of Figure 4A-B indicating neuron ID numbers.

**Supplementary Figure 7.** Brain-wide $Ca^{2+}$ -imaging of basal activity (worm #2).

**Supplementary Figure 8.** Brain-wide $Ca^{2+}$ -imaging of basal activity (worm #3).

**Supplementary Figure 9.** High resolution image of Figure 5A-B indicating neuron ID numbers.

**Supplementary Figure 10.** BAG and URX responses to $O_2$-stimuli in brain-wide WF-TEFO $Ca^{2+}$-imaging.

## Supplementary Table 1

## Supplementary References

**Supplementary Figure 1.** Detectable expression of neurons with *Punc31*- pan-neuronal promoter constructs.

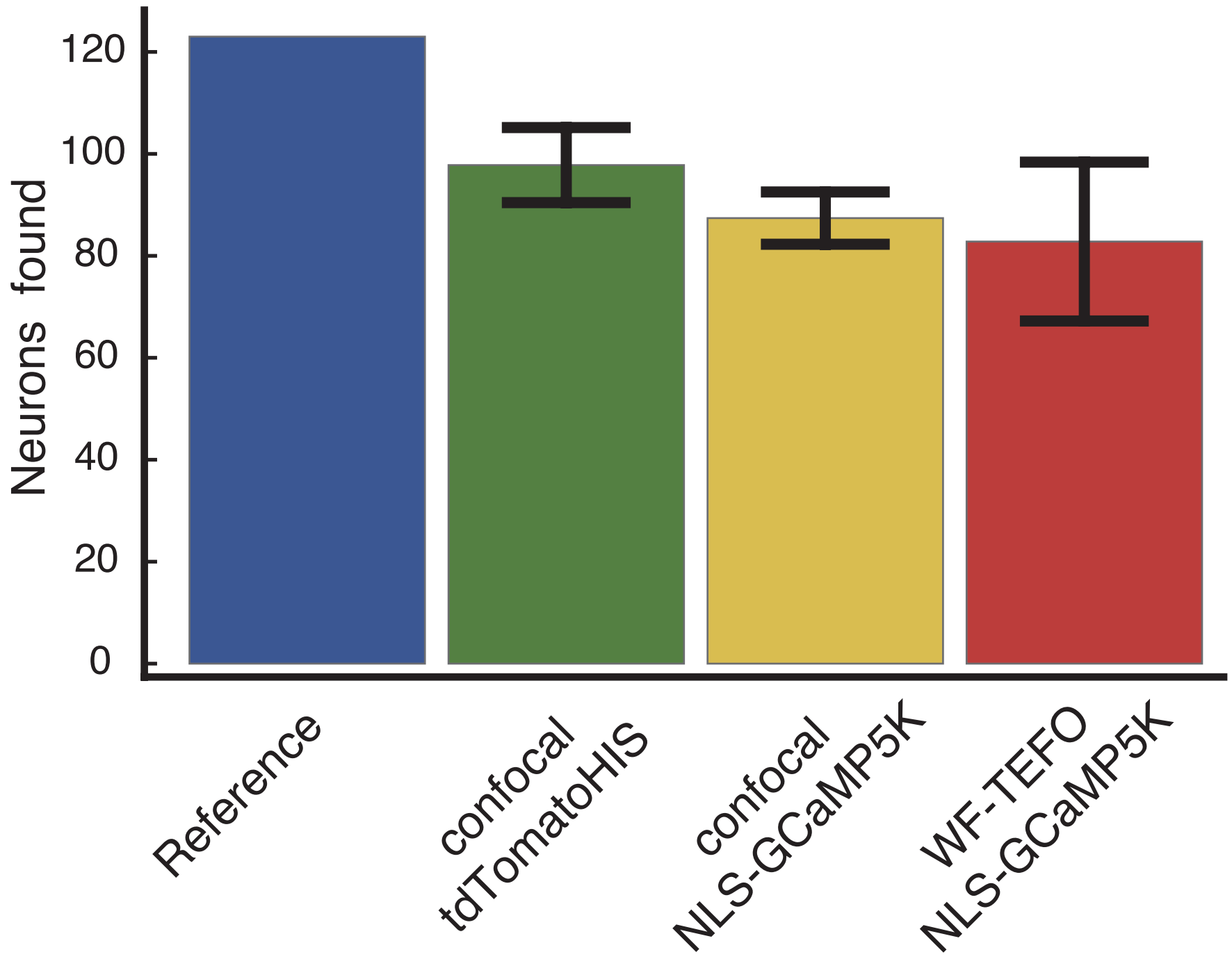

**Supplementary Figure 1. Detectable expression of neurons under *Punc31-* pan-neuronal promoter constructs.**

Here we investigate the ability of the pan-neuronal *Punc31*-promoter to label neurons in the head ganglia region with the indicated expression constructs. The reference (124) is the number of expected neurons localized to our imaging region according to www.wormatlas.org. We collected high-resolution volumetric images using a spinning disk confocal microscope and tested both, worms expressing $Ca^{2+}$-insensitive, nuclear-localized tdTomato-HIS as well as our NLS-GCaMP5K construct, all under the *Punc31*-promoter. We detected 97 ± 8 and 87 ± 5 neurons on average in these cases, respectively. In our high-speed volumetric imaging experiments (WF-TEFO) we are able to record and distinguish 83 ± 15 neurons on average. This implies that we are able to record 70 ± 4% of the neurons in the *C. elegans* head and up to 95% of those that are labeled with our $Ca^{2+}$-reporter. Error bars indicate s.e.m, n = 5 in all cases.

**Supplementary Figure 2.** Behavioral responses to repetitive $O_2$-shifts of wild type and *Punc-31::NLS-GCaMP5K* worms.

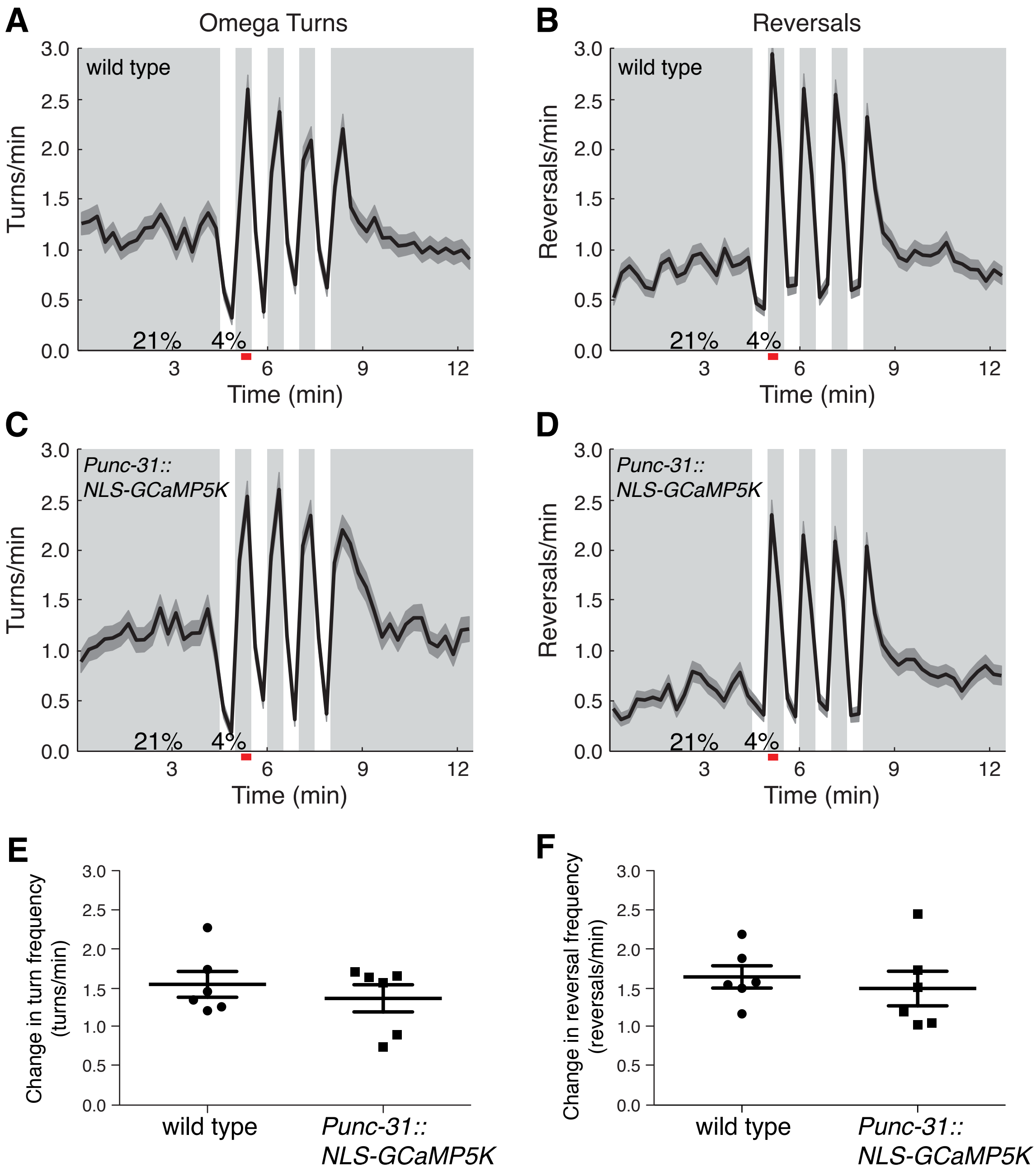

**Supplementary Figure 2. Behavioral responses to repetitive $O_2$-shifts of wild type and *Punc-31::NLS-GCaMP5K* transgenic worms.**

The repetitive $O_2$-shifts protocol (four shifts for 30 s from 21% $O_2$ to 4% $O_2$ and back) used during the $Ca^{2+}$ imaging experiments in Fig. 5 was tested in $O_2$-flow behavioral assays which were performed as described before [13]. **(A-D)** Omega turn and reversal rates of *C. elegans* off food. Omega turns are described as maneuvers that lead to a change in the direction of locomotion after the animals bend their body posture into a Greek-letter-omega like shape. Reversals are brief periods during which the animals move backwards. Traces show averages ± s.e.m of n = 6 equivalent experiments that include 407 (wild type) and 324 (*Punc-31::NLS-GCaMP5K*) worms, respectively. $O_2$ concentrations were switched between 21% and 4%; light shading marks intervals at 21%. **(A, C)** Omega turns per animal per min, calculated in 45 s bins. **(B, D)** Reversals per animal per min, calculated in 45 s bins. **(E,F)** Quantification of changes in turn **(E)** and reversal **(F)** frequency, upon first upshift from 4% to 21% $O_2$ of the 6 independent experiments. Bins for quantification are indicated by red bars in (A-D). Changes are calculated relative to baseline before first downshift. No significant differences were observed between behavioral responses of wild type and *Punc-31::NLS-GCaMP5K* transgenic worms (t-test).

**Supplementary Figure 3.** High resolution image of Figure 3C-D indicating neuron ID numbers.

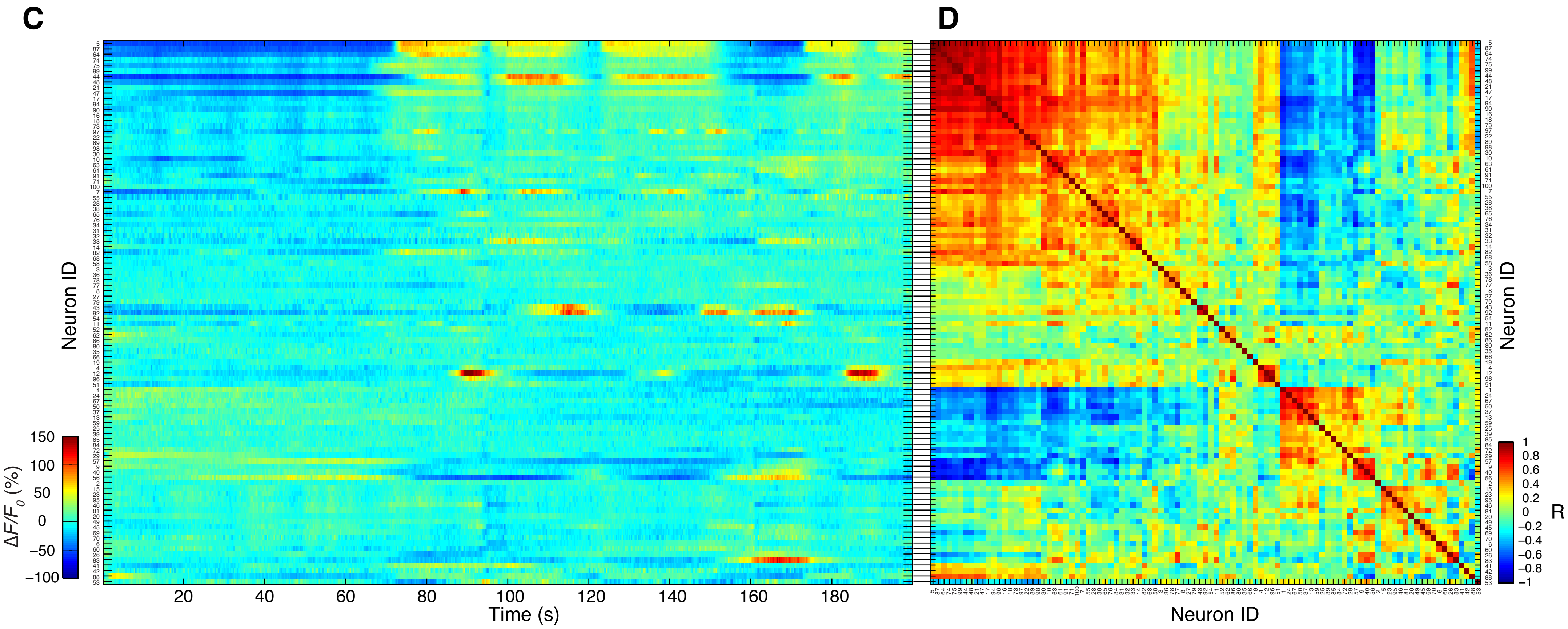

**Supplementary Figure 3. High resolution image of Figure 3C-D indicating neuron ID numbers.**
High resolution representation of Figure 3C-D indicating the neuron ID numbers consistent with numbers in Supplementary Image 1 and Supplementary Data 1.

**Supplementary Figure 4.** Bootstrapping approaches confirming correlation coefficients.

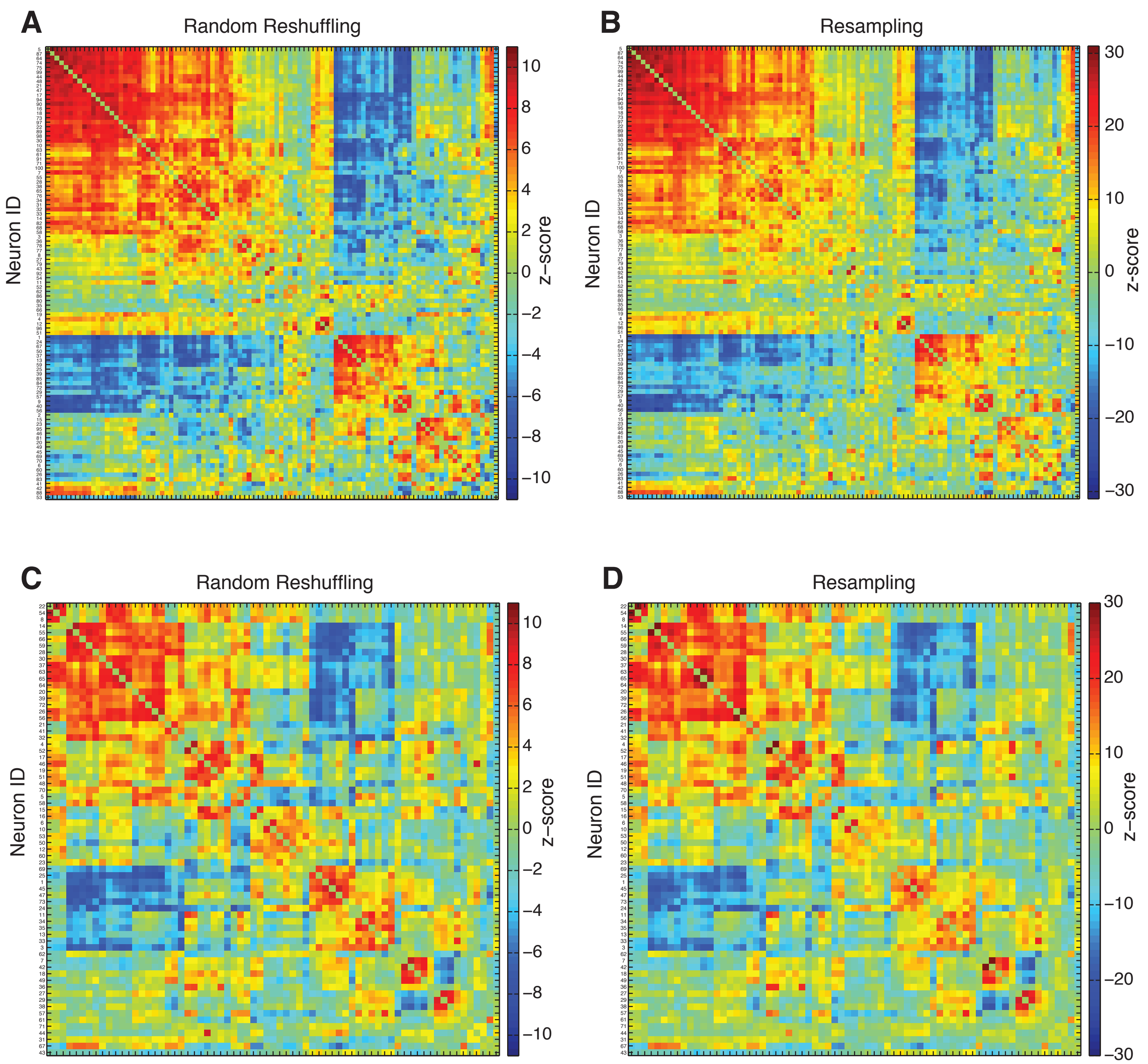

**Supplementary Figure 4. Bootstrapping approaches to confirm correlation coefficients.**

We chose two bootstrapping approaches to confirm that the correlations in our datasets, which are presented in Figs. 3D and 5B, did not arise by chance. **(A)** Correlation matrix of Fig. 3D recalculated from the raw data after random reshuffling. The $\Delta F/F_0$ time series of each neuron has been divided into 10 non-overlapping blocks, randomly reshuffled and the correlation coefficient was recalculated. This was done for each pair of neurons and the z-score of 10,000 such iterations has been plotted. The z-score is defined as $z = (R-\mu)/\sigma$ where $R$ is the measured correlation coefficient, $\mu$ is the bootstrapped mean, and $\sigma$ its standard deviation. It represents the distance between each measured correlation coefficient and the bootstrapped mean in units of the standard deviation. It thus shows the significance of each correlation coefficient. We find that the z-score matrices qualitatively resemble the actually measured matrix of correlations ($R$) that are shown in Fig. 3, which confirms that the correlations presented are indeed real. On average, we find a significance of 8.5 standard deviations across all neuron pairs. **(B)** For comparison, we also present a different bootstrapping approach based on resampling with replacement in which we randomly drew intensities out of the raw data of each neuron until a set of data with equal length to the original was established. This resampled data set was then used for correlation analysis. Again 10,000 iterations of this procedure were used to calculate the z-score of each correlation coefficient that are plotted as a matrix in (B). Again this confirmed the robustness of our high correlation in the original data set as the resampled correlation matrix is qualitatively very similar to one shown in Fig. 3, with an average significance of 3.4 standard deviations across all neuron pairs. **(C-D)** Results for the data used in Fig. 5B. Note that all diagonals have been set to zero as the z-score is undefined for autocorrelations.

**Supplementary Figure 5.** Clustering of data according to covariance.

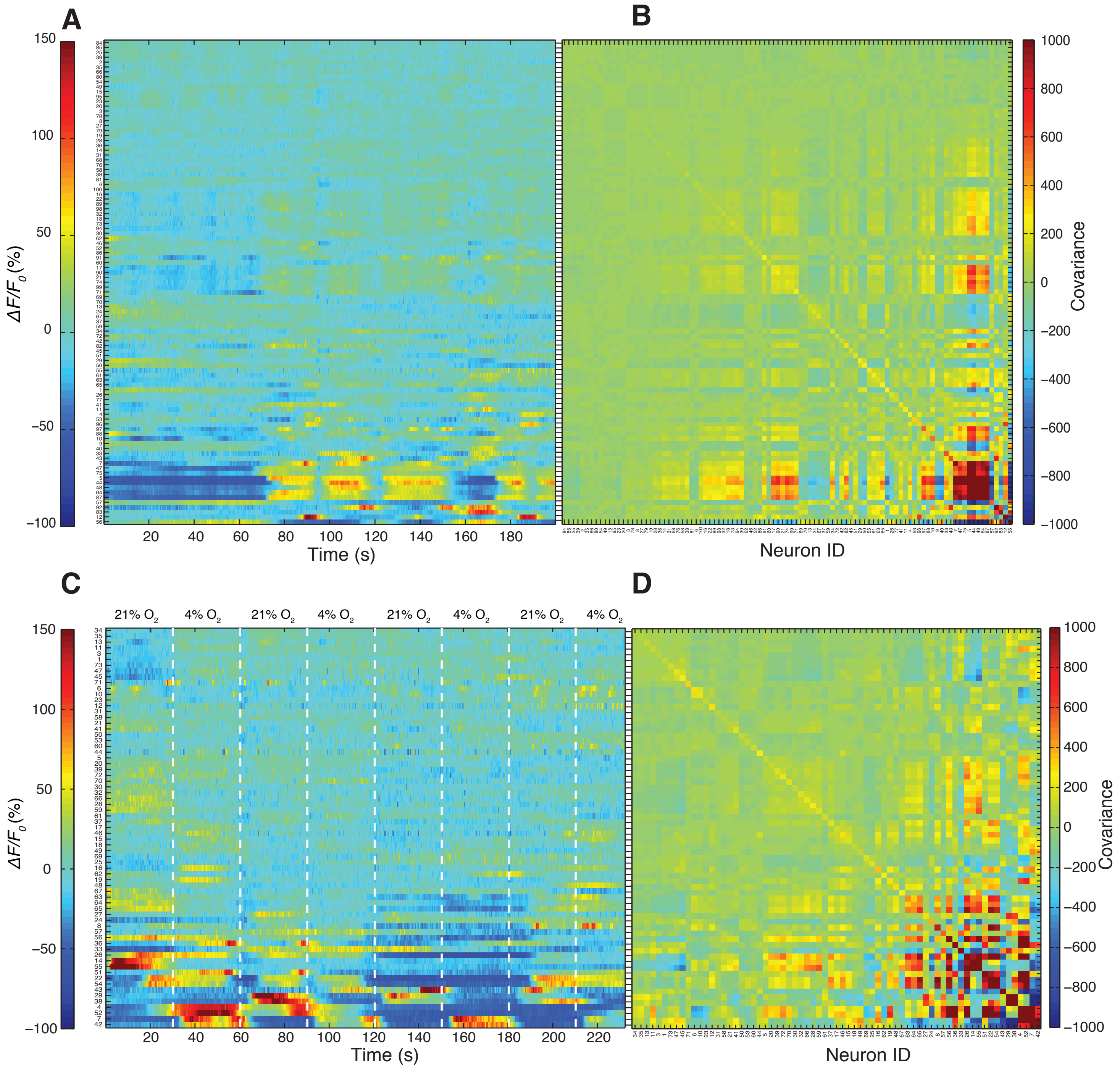

**Supplementary Figure 5. Clustering of data according to covariance.** Here we present an approach to identify co-active neurons in the data set, which is based on covariance. This approach introduces a bias towards larger signal changes, especially in the positive direction, and helped to de-emphasize correlations that arose due to imaging artifacts such as slow movements of inactive neurons across the excitation disk. **(A-B)** Heatplot and covariance matrix for the data presented in Fig. 3C-D. Data where re-ordered based on hierarchical clustering. **(C-D)** Corresponding plots for the data of the stimulated worm in Fig. 5. A manual analysis confirmed that qualitatively this method as well as the one used throughout the main paper identified largely overlapping subsets of active and highly correlated neurons.

**Supplementary Figure 6.** High resolution image of Figure 4A-B indicating neuron ID numbers.

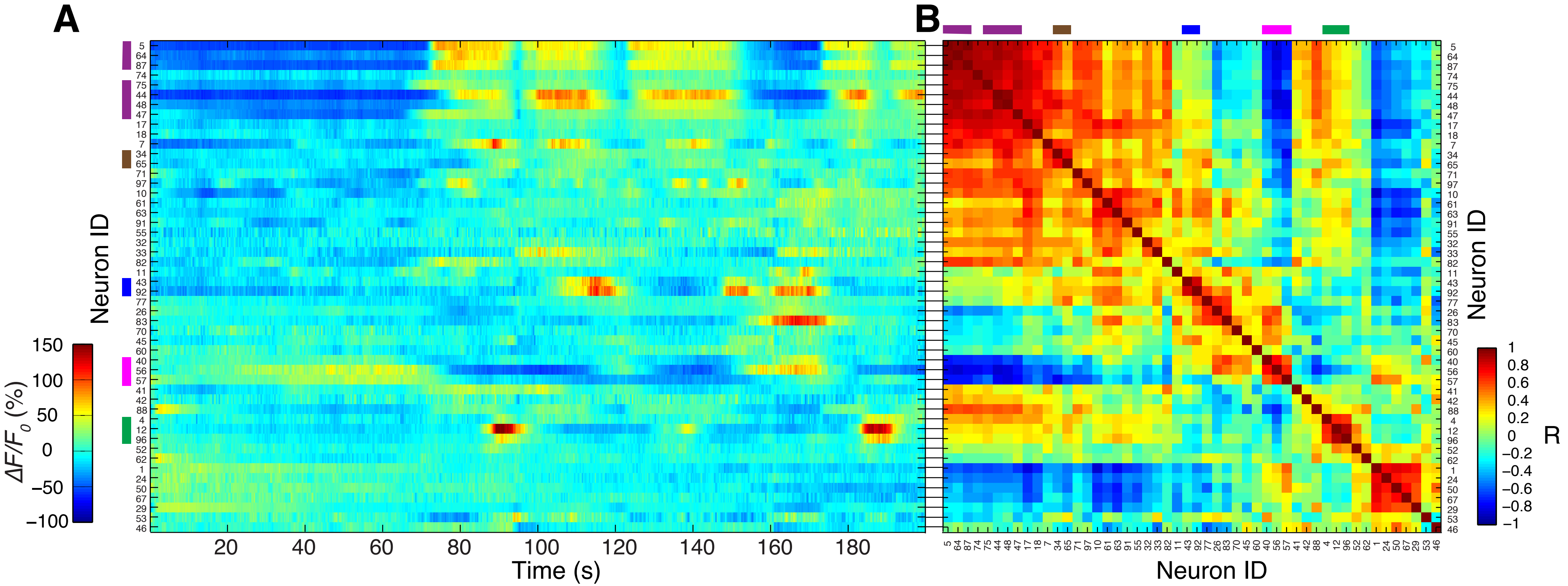

**Supplementary Figure 6. High resolution image of Figure 4A-B indicating neuron ID numbers.**
High resolution representation of Figure 4A-B indicating the neuron ID numbers consistent with numbers in Supplementary Image 1 and Supplementary Data 1.

**Supplementary Figure 7.** Brain-wide $Ca^{2+}$ -imaging of basal activity (worm #2).

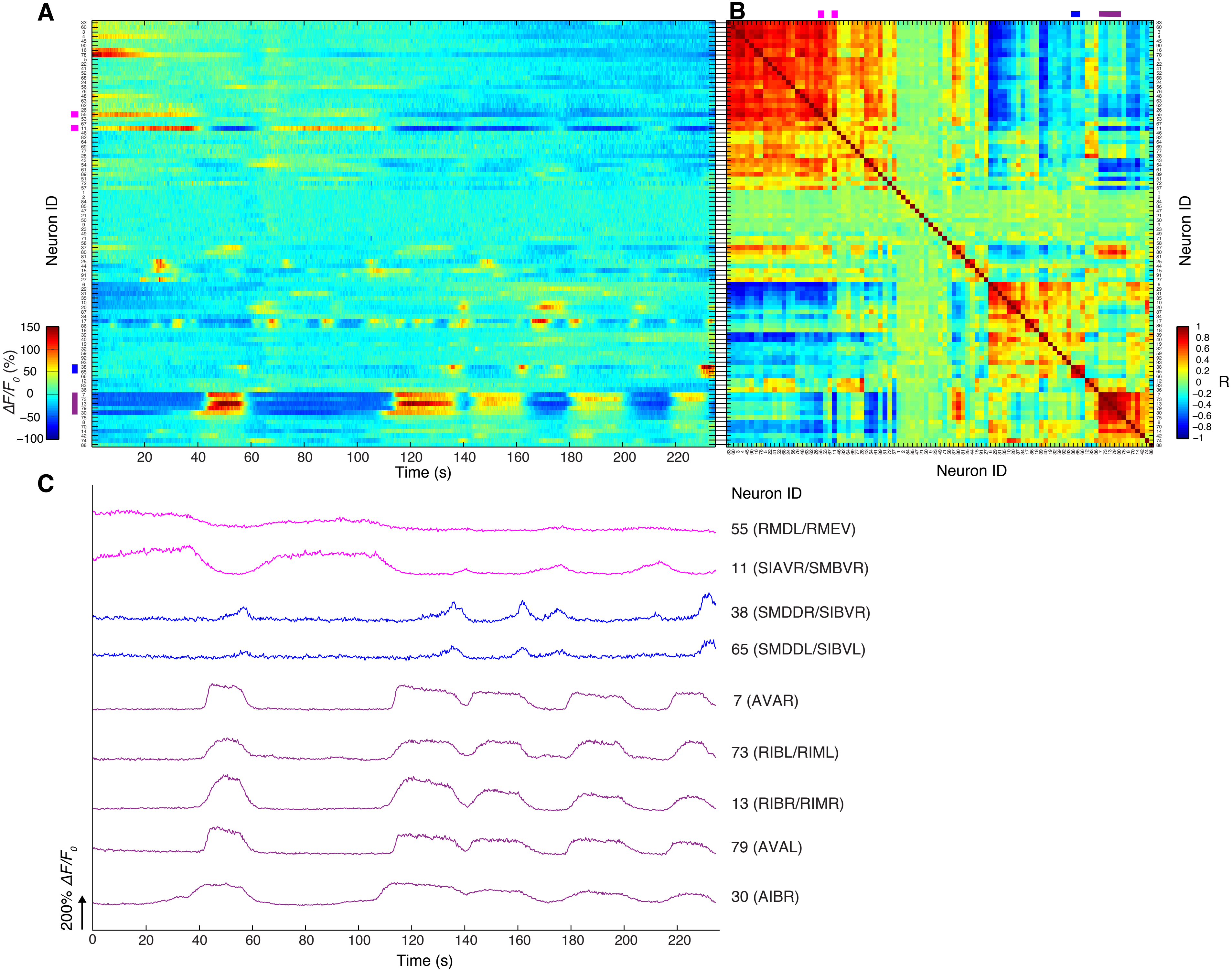

**Supplementary Figure 7. Brain-wide $Ca^{2+}$-imaging of basal activity (worm #2). (A)** Neuronal activity of 93 neurons imaged for 235 s. Each row shows a heat-plot of NLS-GCaMP5K fluorescence time series. Color indicates percent fluorescence changes ($\Delta F/F_0$). $F_0$ for each trace is its mean fluorescence intensity. Colorbar indicates scaling. X-axis represents elapsed recording time. **(B)** Matrix of correlation coefficients ($R$) calculated from the time-series shown in (A). Color indicates the degree of correlation. Colorbar on the right indicates scaling. The data in (A-B) are grouped by agglomerative hierarchical clustering. Corresponding rows in (A) and (B) are aligned. **(C)** Selected traces of neurons. Colors correspond to the clusters indicated by colored boxes in (A-B). Pre-motor interneurons (IDs 7, 73, 13, 79, 30) show 5 synchronized periods of sustained activity, which alternate with neuronal activities whose location is consistent with head-motor neuron classes (IDs 38, 65, 55, 11). Neuronal classes that are consistent with position and morphology of the neurons are indicated next to the neuron ID numbers. All neuron ID numbers are consistent with Supplementary Image 3 and Supplementary Data 3.

**Supplementary Figure 8.** Brain-wide $Ca^{2+}$ -imaging of basal activity (worm #3).

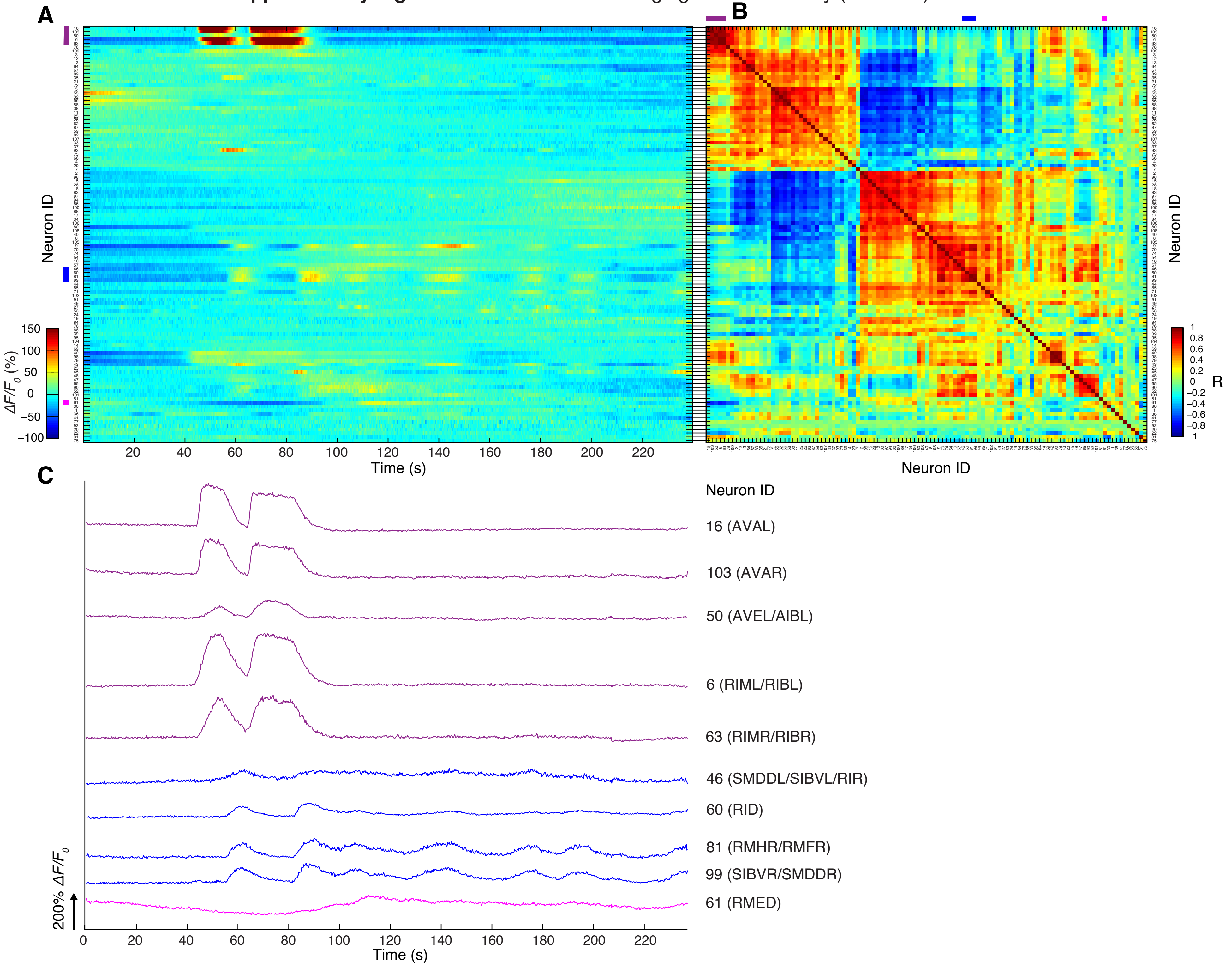

**Supplementary Figure 8. Brain-wide $Ca^{2+}$-imaging of basal activity (worm #3). (A)** Neuronal activity of 109 neurons imaged for 240 s. Each row shows a heat-plot of NLS-GCaMP5K fluorescence time series. Color indicates percent fluorescence changes ($\Delta F/F_0$). $F_0$ for each trace is its mean fluorescence intensity. Colorbar indicates scaling. X-axis represents elapsed recording time. **(B)** Matrix of correlation coefficients (*R*) calculated from the time-series shown in (A). Color indicates the degree of correlation. Colorbar on the right indicates scaling. The data in (A-B) are grouped by agglomerative hierarchical clustering. Corresponding rows in (A) and (B) are aligned. **(C)** Selected traces of neurons. Colors correspond to the clusters indicated by colored boxes in (A-B). Pre-motor interneurons (IDs 16, 103, 50, 6, 63) show 2 synchronized periods of sustained activity, which alternate with neuronal activities whose location is consistent with head-motor neuron classes (IDs 46, 60, 81, 99, 61). Neuronal classes that are consistent with position and morphology of the neurons are indicated next to the neuron ID numbers. All neuron ID numbers are consistent with Supplementary Image 4 and Supplementary Data 4.

**Supplementary Figure 9.** High resolution image of Figure 5A-B indicating neuron ID numbers.

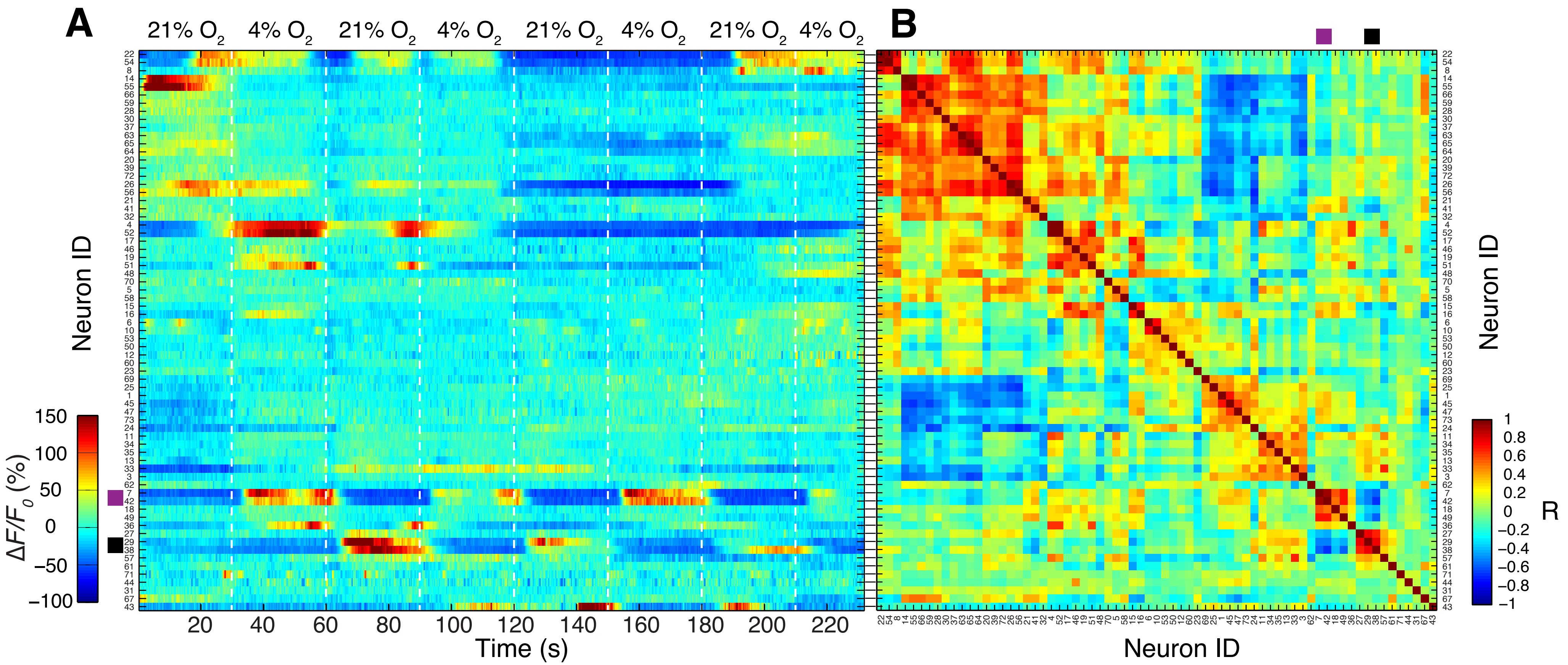

**Supplementary Figure 9. High resolution image of Figure 5A-B indicating neuron ID numbers.**
High resolution representation of Figure 5A-B indicating the neuron ID numbers consistent with numbers in Supplementary Image 2 and Supplementary Data 2.

**Supplementary Figure 10.** BAG and URX responses to $O_2$-stimuli in brain-wide WF-TEFO $Ca^{2+}$-imaging.

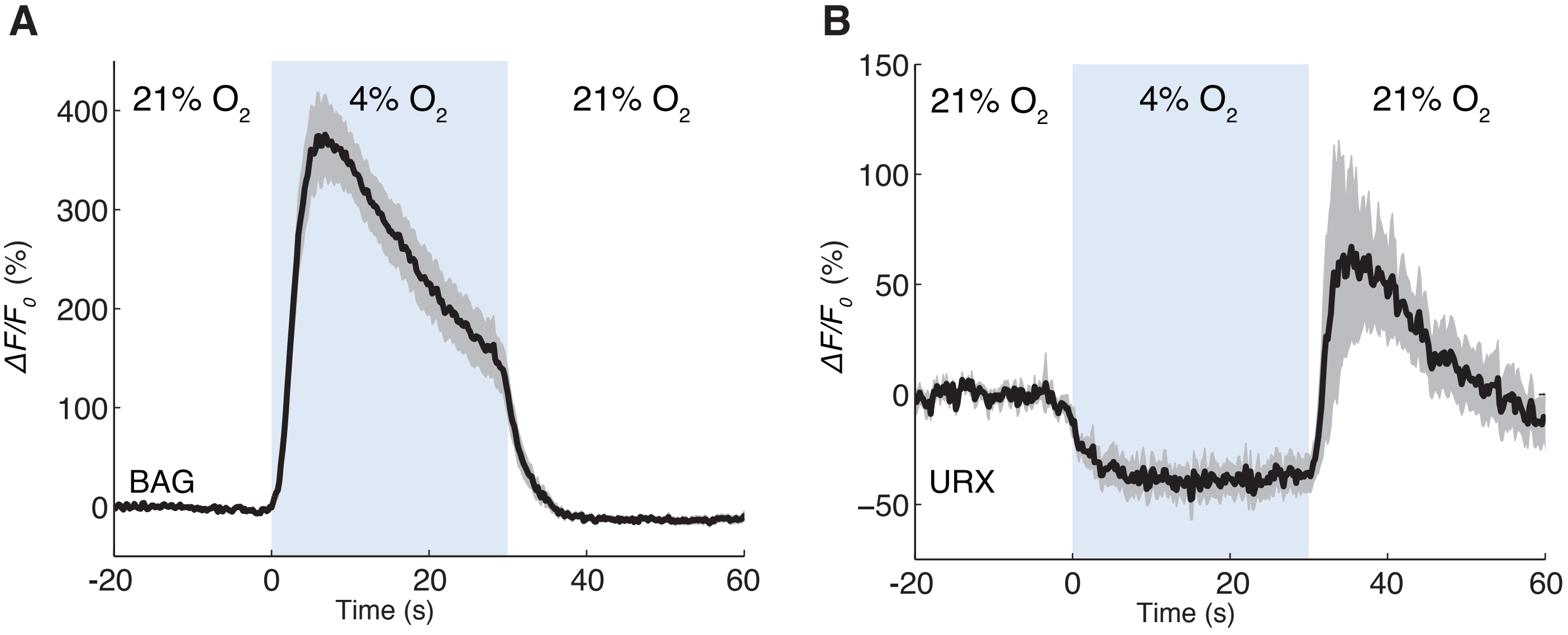

**Supplementary Figure 10. BAG and URX responses to oxygen stimuli in brain-wide WF-TEFO $Ca^{2+}$-imaging. (A-B)** Averaged $Ca^{2+}$-transients of 21 BAG neurons **(A)** and 7 URX neurons **(B)** measured in 11 brain-wide recordings for 230 s. Stimulus protocol included two downshifts from 21% to 4% $O_2$ for 30 s. Responses to first downshift and subsequent upshift are shown. Traces indicate mean $\Delta F/F_0$. Shading indicates s.e.m. X-axis represents elapsed time relative to first $O_2$-downshift. White and blue backgrounds indicate periods at 21% $O_2$ and 4% $O_2$, respectively.

**Supplementary Table 1.** References of pair-wise comparisons of $Ca^{2+}$-imaging experiments and electrophysiological recordings from the same tissues.

| **Tissue** | **Reference $Ca^{2+}$ imaging** | **Reference electrophysiology** |
|---|---|---|
| Pharyngeal muscle | Ref. [1] | Ref. [2] |
| AFD thermosensory neurons | Ref. [3] | Ref. [4] |
| PLM mechanosensory neurons | Ref. [5] | Ref. [6] |
| ASH nociceptive neurons | Ref. [7] | Ref. [8] |
| Pre-motor interneurons | Ref. [8-10] | Ref. [8, 11, 12] |

## Supplementary References